\documentclass[pra,aps,twocolumn,reprint,floatfix,superscriptaddress,longbibliography,amssymb]{revtex4-1}
\usepackage{graphicx}
\usepackage{amsmath,amssymb}
\usepackage{accents}
\usepackage{float}
\usepackage{bbold}

\newcommand{\be}{\begin{equation}}
\newcommand{\ee}{\end{equation}}
\newcommand{\bea}{\begin{eqnarray}}
\newcommand{\eea}{\end{eqnarray}}
\newcommand{\ben}{\begin{equation*}}
\newcommand{\een}{\end{equation*}}
\newcommand{\ba}{\begin{align}}
\newcommand{\ea}{\end{align}}

\newcommand{\mbf}{\mathbf}
\newcommand{\mrm}{\mathrm}

\renewcommand{\vec}[1]{\mathbf{#1}}

\begin{document}

\title{Emergent and broken symmetries of atomic self-organization \\ arising from Gouy phase shifts in multimode cavity QED}

\author{Yudan Guo}
\affiliation{Department of Physics, Stanford University, Stanford, CA 94305}
\affiliation{E.~L.~Ginzton Laboratory, Stanford University, Stanford, CA 94305}
\author{Varun D.~Vaidya}
\affiliation{Department of Physics, Stanford University, Stanford, CA 94305}
\affiliation{E.~L.~Ginzton Laboratory, Stanford University, Stanford, CA 94305}
\affiliation{Department of Applied Physics, Stanford University, Stanford, CA 94305}
\author{Ronen M.~Kroeze}
\affiliation{Department of Physics, Stanford University, Stanford, CA 94305}
\affiliation{E.~L.~Ginzton Laboratory, Stanford University, Stanford, CA 94305}
\author{Rhiannon A.~Lunney}
\altaffiliation[Present address: ]{School of Physics and Astronomy, Cardiff University, Cardiff CF24 3AA, United Kingdom}
\affiliation{SUPA, School of Physics and Astronomy, University of St Andrews, St Andrews KY16 9SS UK}
\author{\\Benjamin L.~Lev}
\affiliation{Department of Physics, Stanford University, Stanford, CA 94305}
\affiliation{E.~L.~Ginzton Laboratory, Stanford University, Stanford, CA 94305}
\affiliation{Department of Applied Physics, Stanford University, Stanford, CA 94305}
\author{Jonathan Keeling}
\affiliation{SUPA, School of Physics and Astronomy, University of St Andrews, St Andrews KY16 9SS UK}

\date{\today}

\begin{abstract}
Optical cavities can induce photon-mediated interactions among intracavity-trapped atoms.  Multimode cavities provide the ability to tune the form of these interactions, e.g., by inducing a nonlocal, sign-changing term to the interaction.   By accounting for the Gouy phase shifts of the modes in a nearly degenerate, confocal, Fabry-P\'{e}rot cavity, we provide  a theoretical description of this interaction,  along with additional experimental confirmation to complement that presented in the companion paper, Ref.~\cite{GouyPRL2018}.  Furthermore, we show that this interaction should be written in terms of a complex order parameter, allowing for a $\mathbb{U}(1)$-symmetry to emerge.  This symmetry corresponds to the phase of the atomic density wave arising from self-organization when the cavity is transversely pumped above a critical threshold power.  We theoretically and experimentally show how this phase depends on the  position of the Bose-Einstein condensate (BEC) within the cavity and discuss mechanisms that break the $\mathbb{U}(1)$-symmetry and lock this phase.   We then  consider  alternative Fabry-P\'{e}rot multimode cavity geometries (i.e., beyond the confocal) and schemes with more than one pump laser and show that these provide additional capabilities for tuning the cavity-meditated interaction among atoms, including the ability to restore the $\mathbb{U}(1)$-symmetry despite the presence of symmetry-breaking effects. These photon-mediated interactions may be exploited for realizing quantum liquid crystalline states and  spin glasses using multimode optical cavities.
\end{abstract}

\maketitle

\section{Introduction}
The cavity QED system of quantum degenerate atoms trapped inside an optical cavity opens numerous possibilities for studying quantum many-body systems in a driven-dissipative setting~\cite{Ritsch2013}. Recent advances include demonstration of supersolidity~\cite{Leonard2017}, supermode-polariton condensation~\cite{Kollar2017}, and spinor self-ordering~\cite{kroeze2018spinor}. Beyond the traditional single-mode cavity system, degenerate multimode cavities have emerged as promising platforms for realizing nontrivial interactions among trapped atoms~\cite{VaidyaPRX} and exotic photonic matter~\cite{Schine2016}.  Simultaneously addressing the multiple degenerate (or nearly degenerate) modes of a cavity, which may have distinct and incommensurate transverse spatial profiles,  greatly expands the number of transverse degrees of the freedom in the cavity.  This, in turn, provides the ability to engineer tunable-range cavity-photon-mediated interactions, as described in our previous work of Ref.~\cite{VaidyaPRX}. 

A less-studied and subtle point is the role of the differing longitudinal profiles of these degenerate modes, particularly the effect of the Gouy phase shift contribution~\cite{Gouy:1890,*Gouy:1891}\footnote{The many interpretations of the Gouy phase shift are discussed in Refs.~\cite{Boyd:1980ek,siegman1986lasers,Feng*2001,Padgett:2008bi,Visser:2010dx}.}. As briefly mentioned in Ref.~\cite{VaidyaPRX}, and presented more extensively in the companion paper of Ref.~\cite{GouyPRL2018}, the Gouy phase shifts affect how different transverse modes sum to produce an effective nonlocal and sign-changing cavity-mediated interaction in the transverse direction.  The present work explores the origin of this effect and its consequences. 
 
For a BEC trapped in a transversely pumped cavity, atoms will start to self-organize coinciding with the superradiant emission of cavity photons; i.e., atoms will form a density wave  commensurate with the optical lattice due to the dynamically generated cavity light when the cavity-mediated interaction overcomes the kinetic energy cost of the associated density wave~\cite{Domokos2002,Black2003,Baumann2010,Ritsch2013,Kollar2017,VaidyaPRX}.  Compared to the case of a single-mode cavity~\cite{Baumann2010}, where the density wave is forced to conform to the shape of the single cavity mode and the transition is adequately described by the Hepp-Lieb-Dicke model~\cite{Ritsch2013,Kirton:2018vv}, atoms in a degenerate cavity have more freedom. The interference of many transverse modes allows the atoms to 
adopt a far wider range of shapes, which can induce a more exotic phase transition~\cite{Gopalakrishnan2009,Gopalakrishnan2010}. Moreover, the intracavity photons can be in a superposition of many modes, leading to a localized photonic wavepacket. This induces a localized, short-range photon-mediated interaction between atoms~\cite{VaidyaPRX}. In addition, the distinct longitudinal profiles of the transverse modes, due to Gouy phase shifts, result in a phase degree of  freedom of the  density wave. That is, not only the amplitude, but also the phase of the atomic density wave can vary across the cavity and can in principle become free (in the sense of a $\mathbb{U}(1)$-symmetry) under special conditions. 
These features, especially the enhanced phase space of the order parameter, greatly enrich the physics of the self-organization transition.

In this paper, we present a detailed theoretical and experimental study of how Gouy phase shifts induce this nonlocal interaction and determine the atomic density wave phase degree of freedom through this interaction.
Section~\ref{cavmedint} presents a general derivation of cavity-mediated interactions, complementing that first discussed in our Ref.~\cite{VaidyaPRX}, while also deriving a general form of the nonlocal interaction studied in the companion paper~\cite{GouyPRL2018}.
Section~\ref{selforg} then discusses how these interactions determine the density wave state arising from the Dicke superradiant, self-ordering density-wave transition studied in the companion paper~\cite{GouyPRL2018}. We also show how the spatial structure of the field emitted from the cavity reveals important information about the phase of the atomic density wave.  Restricting to an ideal confocal cavity, we show that the atomic density wave possesses a continuous degree of phase freedom and a $\mathbb{U}(1)$-symmetry emerges at special positions in the cavity. Section~\ref{U1sym} discusses this $\mathbb{U}(1)$-symmetry in an ideal confocal cavity before Sec.~\ref{breaking-u1} describes the  mechanisms that break this symmetry in realistic confocal cavities coupled to BECs of finite size. Results of experiments that image cavity field emission  are presented, demonstrating this symmetry breaking. 
Section~\ref{extensions} then describes the consequences of this symmetry breaking for realizing various quantum many-body systems such as quantum liquid crystalline states~\cite{Gopalakrishnan2009, Gopalakrishnan2010}.  Section~\ref{restoration} presents a proposal for a pumping scheme that restores the $\mathbb{U}(1)$-symmetry and is robust against various possible symmetry breaking effects. Generalization of the cavity mediated interaction to three dimensions is also discussed in Sec.~\ref{threeDimensions}. Lastly, before concluding remarks, we derive in Sec.~\ref{beyond-confocal} the form that cavity-induced interactions take in Fabry-P\'{e}rot multimode cavities with geometries beyond the confocal. 

\section{Cavity-mediated interaction and self-organization}
\label{cavmedint}

We start by introducing a model of $N$ atoms in a Bose-Einstein condensate (BEC) with wavefunction $\Psi(\vec{x})$ and interacting with cavity modes described by annihilation operators $\hat{a}_{\mu,Q}$.  The cavity axis is taken to lie along $\hat{z}$, while the transverse, standing-wave pump field is oriented along $\hat{x}$. The Hamiltonian for this can be written as:
\begin{multline}
\label{eq:Hamiltonian}
H= -\sum_{\mu,Q}\Delta_{\mu,Q} \hat{a}^\dagger_{\mu,Q} \hat{a}^{}_{\mu,Q}  \\
+ N\int d^3\mbf{x}
\Psi^*(\mbf{x})\left(-\frac{\nabla^2}{2m}+ V(\mbf{x}) 
+\frac{U}{2}|\Psi(\mbf{x})|^2\right)\Psi(\mbf{x})\\
+\frac{N}{\Delta_a} \int d^3\vec{x}
\Psi^*(\mbf{x})|\hat{\phi}|^2
\Psi(\mbf{x}).
\end{multline}
For compactness, we use $\mu = (l,m)$ to index the transverse modes of our cavity (in the Hermite-Gaussian basis of transverse electromagnetic modes TEM$_{\mu}$), and we  define the total mode family index $n_\mu = l_\mu + m_\mu$. The index $Q$ defines the longitudinal mode number;  we keep the dependence on this explicit for now. When we later consider a given family of degenerate modes in a confocal or equivalent cavity, the longitudinal and transverse mode numbers are related to one another. This behavior is discussed in detail for general cases of multimode cavity geometries in Sec.\ref{beyond-confocal}.

The first term in Eq.~\eqref{eq:Hamiltonian} is the Hamiltonian of the cavity modes with detuning from the transverse pump field $\Delta_{\mu,Q}$.  The second line is the standard Hamiltonian for a weakly interacting BEC with contact interactions of strength $U$ in an external trap $V(\vec{x})$.  The final line is the optical potential from the Stark shift, proportional to $1/\Delta_a$, due to the combined cavity and pump light. The light field $\hat{\phi}$ thus consists of the standing-wave pump and a sum over all cavity modes with their transverse and longitudinal spatial dependence~\cite{siegman1986lasers},
\begin{multline}
\hat{\phi}(\mbf{r}) = \Omega \cos(k_rx) \\+ g_0\sum_{\mu,Q} \hat{a}_{\mu,Q}
\Xi_\mu(\mbf{r})\cos{\left[k_r\left(z+\frac{r^2}{R(z)}\right)-\theta_{\mu,Q}(z)\right]}.
\label{totalLightField}
\end{multline}
Here, $\Omega$ is the pump's Rabi frequency, $k_r$ is the recoil momentum of the atom, $g_0$ is the bare atom-cavity coupling strength defined for an atom coupled to the peak intracavity field of the TEM$_{00}$ mode, and $\Xi_\mu(\vec{r})$ are Hermite-Gauss mode functions of the cavity.  In writing in Eq.~\eqref{totalLightField}, we have included the radial dependence of the phase $r^2/R(z)$, where $R(z)=z+z_R^2/z$ is the radius of curvature of the phase fronts at longitudinal position $z$ and $z_R$ is the Rayleigh range. When we consider atoms near the cavity midplane $z=0$, we have that $R(z) \to \infty$, so the phase fronts become flat and this contribution to the phase can be neglected. The form of Eq.~\eqref{totalLightField} results in a spatially varying single-photon Rabi frequency $g_0\Xi_\mu(\vec{r})/\Xi_{00}(0)$ for the mode $\mu$ (assuming atoms are located at the antinodal planes along the cavity). Note that here, and in the following, we use $\mbf{r}$ to denote the coordinate in the plane transverse to the cavity axis. 

The cosine appearing in Eq.~\eqref{totalLightField} also contains the term $\theta_{\mu,Q}(z)$, which describes terms varying slowly compared to $k_rz$. The subscripts indicate that this term depends both on the transverse index $\mu$ and the longitudinal mode number $Q$.  We
may write this term as~\cite{siegman1986lasers}:
\begin{equation}
  \theta_{\mu,Q}(z) = \psi(z) + n_\mu [\psi(L/2) + \psi(z)] - \xi_{\mu,Q}.
  \label{eq:lightfieldphase}        
\end{equation}
This term accounts for several effects.  First, in order to satisfy Maxwell's equations, terms which vary faster in the transverse plane must vary more slowly along the cavity axis.  Thus, the $z$ dependence of the phase  is given by $\psi(z)=\arctan(z/z_R)$. The existence of this extra phase evolution, known as the Gouy phase shift, or phase anomaly~\cite{Gouy:1890,*Gouy:1891,siegman1986lasers}, and particularly its $n_\mu$ dependence (i.e., the phase shift increases for greater $n_\mu$), is the source of the rich physics discussed in this paper.
The other terms in Eq.~\eqref{eq:lightfieldphase}, $n_\mu \psi(L/2)$ and $\xi_{\mu,Q}$, are phase offsets required to match boundary conditions at the mirrors (which are placed at $z=\pm L/2$).

As noted earlier, atoms subject to transverse pumping, as described by Eq.~\eqref{eq:Hamiltonian}, will undergo self-organization to form an atomic density wave commensurate with the optical lattice due to the pump and cavity light.  To describe this transition, we need to write the motional wavefunction of the atoms, allowing for scattering of atoms by the combined pump and cavity lattices.
We assume that most of the condensate is in the ground state, with a small fraction having undergone a momentum kick from either scattering a photon from the retroreflected transverse pump into the cavity or vice-versa. Hence, we write
\begin{multline}
\label{atomwf}
\Psi(\mbf{x})=Z(z-z_0) \sqrt{\rho(\mbf{r})}\big[\psi_0 + \sqrt{2}\cos(k_rx) \times \\ \left\{\psi_c \cos(k_rz + \delta)+\psi_s \sin(k_rz +\delta)\right\}\big],
\end{multline}
where $Z(z)$ is an envelope function which describes the confinement of the gas at a position $z_0$ along the $\hat{z}$ direction, $\rho(\mbf{r})$ is the density profile of the atoms in the cavity transverse plane, $\psi_0$ is the amplitude of the ground-state condensate wavefunction of the gas, and $\psi_{c,s}$ are the amplitudes of the parts of the gas that have been scattered into two out-of-phase density profiles.  These have an undetermined phase offset $\delta$.  We have written two independent components $\psi_c, \psi_s$ to allow us to describe an emergent freedom between the amplitudes of these two components. The phase offset $\delta$  is arbitrary, and in the following we will choose a value which enables us to simplify subsequent expressions.  We may also consider $\psi_c$ and $\psi_s$ as being the real and imaginary components of a complex order parameter describing the atomic density wave, i.e., using $\psi_1=\psi_c + i \psi_s$ and thus writing: 
$$\Psi(\mbf{x})=Z(z-z_0) \sqrt{\rho(\mbf{r})}[\psi_0 + \sqrt{2}\cos(k_rx)\Re\{ \psi_1 e^{i k_r z + \delta}\}].$$

We can now use Eq.~\eqref{eq:Hamiltonian} to find the mean-field equations of motion for $\psi_{0,c,s}$ and $\alpha_{{\mu,Q}}
\equiv\langle\hat{a}_{{\mu,Q}}\rangle$. It is convenient to write equations in terms of only the transverse coordinates
$\mbf{r}$, for which we must perform the $z$  integral in Eq.~\eqref{eq:Hamiltonian}.  This can be done straightforwardly in the limit where we assume $Z(z)$ has a width $w_z$ and that $\lambda\ll w_z\ll z_R$. The first inequality allows us to drop any terms oscillating at wavevector $k_r$ or $2k_r$; this imposes momentum conservation so that recoiling atoms pick up the difference of pump and cavity momenta.  The second condition means that we can evaluate the slowly varying phase terms as being effectively constant over the width of the gas:  we can approximate $\theta_{\mu,Q}(z) \simeq \theta_{\mu,Q}(z_0)$.  Both conditions are well-satisfied in typical experiments~\cite{VaidyaPRX}.

In this paper, we will consider the onset of self-organization, where the $\psi_{c,s}$ become non-zero, leading to an atomic density wave with an associated occupation of the cavity modes.  To understand the transition to this state, and the patterns of atomic density waves which become occupied, we focus on linear stability of the normal state.  That is, we linearize in the variables $\psi_c,\psi_s$ and $\alpha_\mu$.  In this linearized treatment, we find that all the relevant $z$ integrals involve a cross term between pump light and cavity light that causes scattering between at-rest atoms $\psi_0$ and $\psi_{c,s}$. We then find that the $z$ integrals yield two possible overlap values,
\begin{equation}
\label{generaloverlapfator}
\mathcal{O}_{\mu,Q}^\sigma=\begin{cases} \cos[\theta_{\mu,Q}(z_0) - \delta] & \sigma=c \\ \sin[\theta_{\mu,Q}(z_0) - \delta] & \sigma=s\end{cases}, 
\end{equation}
where the superscript $\sigma$ distinguishes between two possible out-of-phase density waves $\psi_{c,s}$.  In Sec.~\ref{beyond-confocal}, we will revisit the assumption that we may approximate the $z$ dependence of $\theta_{\mu,Q}$ by its value at $z=z_0$ and consider the leading order corrections to this result.

In the linearized regime, we need only consider equations for $\psi_\sigma$ and $\alpha_{\mu,Q}$ because the ground-state amplitude $\psi_0$ can be considered constant. Using the above overlaps, we find the linearized equations take the form:
\begin{widetext}
\begin{align}
\label{eq:eom}
&i\partial_t\alpha_{\mu,Q} =-(\Delta_{\mu,Q}+i\kappa)\alpha_{\mu,Q} -\frac{g_0^2 N}{2\Delta_a}
\int d\mbf{r} |\psi_0|^2 \rho(\mbf{r})\Xi_\mu(\mbf{r})\Xi_\nu(\mbf{r}) \alpha_{\nu,Q}
-\frac{g_0 N\Omega}{2\sqrt{2}\Delta_a}\int d\mbf{r}\Xi_\mu(\mbf{r}) \rho(\mbf{r})
\displaystyle\sum_{\sigma=c,s}\left[\psi_0^{\ast}\psi_{\sigma}^{}+\psi_0^{}\psi^{\ast}_{\sigma}\right]\mathcal{O}_{\mu,Q}^\sigma \\
\label{eq:eom2}
&i\partial_t\psi_{\sigma} =\int d\mbf{r} \left(\mu+2\omega_r \right) \rho(\mbf{r}) \psi_{\sigma}+\frac{1}{2}\int d\mbf{r} U\psi_0^{2}\psi^{\ast}_{\sigma}\rho(\mbf{r})-\frac{g_0\Omega}{2\sqrt{2}\Delta_a}\int d \mbf{r} \sum_{\mu,Q}(\alpha_{\mu,Q}^\ast+\alpha_{{\mu,Q}})\Xi_\mu(\mbf{r})\psi_0 \rho(\mbf{r}) \mathcal{O}_{\mu,Q}^\sigma,
\end{align}
\end{widetext}
where we have included the photon loss rate $2\kappa$, $\mu$  is the chemical potential of the ground-state condensate (when not used as an index), and $\omega_r$ is the recoil energy $k_r^2/2m$.

Since we expect the cavity field to reach a steady state on a timescale much faster than the atomic motion, we  adiabatically eliminate the photons by setting the time derivative in Eq.~\eqref{eq:eom} to zero and solve for $\alpha_{\mu,Q}$. We also neglect the corrections to the bare cavity modes caused by the ground-state atomic gas; i.e., the term proportional to the integral of $|\psi_0|^2$ is neglected. Substituting $\alpha_{\mu,Q}$ back into the equation of motion of the atomic condensate gives
\begin{multline}
\label{eq:atomeom}
i\partial_t\psi_{\sigma}=\int d\mbf{r} \left(\mu+2\omega_r \right)
 \rho(\mbf{r}) \psi_{\sigma}+\frac{1}{2}\int d\mbf{r} U\psi_0^{2}\psi^{\ast}_{\sigma}\rho(\mbf{r}) \\
+\frac{g_0^2 \Omega^2 N}{2 \Delta_a^2\Delta_{\mu_0,Q_0}}\int d\mbf{r} \int d\mbf{r}^\prime \sum_{\tau=c,s}
\Re\{\mathcal{D}_{\sigma \tau}(\mbf{r},\mbf{r}^\prime)\}
\rho(\mbf{r})\rho(\mbf{r}^\prime) \\ 
 \times|\psi_0|^2\Bigl[\psi_{\tau}
+\psi^{\ast}_{\tau}\Bigr],
\end{multline}
where $\Delta_{\mu_0,Q_0}$ is taken as the detuning of some reference mode (which, later we will take as the fundamental mode in a degenerate family), and the cavity-mediated interaction takes the form:
\begin{equation}
\label{eq:kernel}
\mathcal{D}_{\sigma \tau}(\mbf{r},\mbf{r}^\prime)=\Delta_{\mu_0,Q_0}\sum_{{\mu,Q}}\frac{\Xi_{\mu}(\mbf{r})\Xi_{\mu}(\mbf{r}^\prime)}{\Delta_{{\mu,Q}}+i\kappa}\mathcal{O}_{{\mu,Q}}^\sigma \mathcal{O}_{{\mu,Q}}^{\tau}.
\end{equation}
This interaction matrix describes the interaction in terms of the real and imaginary components $\psi_\tau$  of the complex order parameter describing atomic density waves, with freedom of both density wave amplitude and phase.

Expression~\eqref{eq:kernel} captures the general cavity-mediated interaction between the two density wave components without any assumptions about the structure of the cavity modes or their degeneracies. In the following sections, we will focus on a confocal cavity, where many modes are degenerate (or nearly so in realistic cavities), and restrict our attention to that family of nearly resonant modes closest to the pump frequency, with all other mode families being relatively far detuned.  

\section{Cavity-mediated interactions in a confocal cavity}\label{selforg}

In this section, we consider how the general results of the previous
section apply for a confocal cavity.  As noted in the introduction and previous section, for a degenerate cavity, the existence of many degenerate modes allows for both transverse spatial variation of the light and for transverse variation of whether it couples to sine or cosine atomic density waves.  To explore this, we will examine how the effective cavity-mediated interaction matrix behaves in the confocal limit.

We first summarize the standard results~\cite{siegman1986lasers} for the parameters defining the mode functions in a confocal cavity; a general derivation of equivalent results for all degenerate Fabry-P\'{e}rot cavities is given in Sec.~\ref{beyond-confocal}.
A confocal cavity is defined by the cavity geometry $R = L$, where $R$ is the radius of curvature of the mirrors and $L$ is the length of the cavity. Matching the curvature of phase fronts with the radius of curvature of the mirrors, we find that the Rayleigh range $z_R = L/2$.  For a given family of degenerate modes in such a cavity, the longitudinal mode number
$Q$ is locked to $\mu$ via $Q=Q_0-(n_\mu-\eta)/2$, where $Q_0$ determines the longitudinal mode number of the lowest transverse mode within a given degenerate family and $\eta=0$ or 1.   As expected for a confocal cavity, this formula leads to separately resonant families of odd and even transverse modes as selected by $\eta$. We consider a regime where one specific degenerate family is near-resonant with the pump laser and all other degenerate frequencies can be neglected. Because of this,  we can suppress the $Q$ dependence of the formulae written in Sec.~\ref{cavmedint}, and sum only over transverse modes.  We can however allow small deviations from confocality, by considering the detunings within the near-resonant family to take the form $\Delta_{\mu, Q} \to \Delta_\mu = \Delta_{Q_0} + n_\mu \epsilon$  where $\Delta_{Q_0}$ is the detuning of the fundamental mode in a given family and  $\epsilon$ describes the residual splitting of near-degenerate mode frequencies \cite{VaidyaPRX}.

With this restriction to a single degenerate family, we can also simplify the overlap factors.  Because $Q$ and $\mu$ are locked, we find that within a given degenerate family, the phase shift $\xi_{\mu,Q}$ required to match boundary conditions becomes constant, i.e., $\xi_{\mu,Q}=\xi_{Q_0}$.  The slow phase dependence can therefore be written as $\theta_{\mu,Q}(z_0) = \Theta_{Q_0}(z_0) + n_\mu \theta_0(z_0)$, where we 
have explicitly that
\be
\theta_0(z_0)=\pi/4 + \mrm{arctan}(z_0/z_R).
\label{zphase}
\ee
and
 $\Theta_{Q_0}(z)=\psi(z) - \xi_{Q_0}$.  We may also note that $\xi_{Q_0}=(\pi/2)(Q_0+1)$ for even parity modes, so the longitudinal profiles of the modes one free-spectral-range (FSR) apart are phase shifted by a quarter period.
Furthermore, we may now choose the phase offsets of the atomic density waves as $\delta=\Theta_{Q_0}$, simplifying the overlap factors as
\begin{equation}
\label{overlapfator}
\mathcal{O}_{\mu,Q}^\sigma=\begin{cases} 
\cos(n_\mu \theta_0) & \sigma=c \\ 
\sin(n_\mu \theta_0) & \sigma=s
\end{cases}. 
\end{equation}
For brevity, we have suppressed the $z_0$ dependence of $\theta_0$ here and in the following expressions.

In this confocal limit, where sums are restricted to modes within a given family, we may simplify the interaction matrix in  Eq.~\eqref{eq:kernel} significantly.  
Since, as already noted, a confocal cavity only supports degenerate families with the same parity, to find the cavity-mediated interaction, summation over transverse modes should be restricted to $n_\mu$ being either even or odd. This can be done by introducing an appropriate factor $\mathcal{S}_{\mu}$ in Eq.~\eqref{eq:kernel} to cancel unwanted modes in a particular confocal cavity family; the sum is then made to be over all transverse modes.  This factor  should be chosen as
\be
\mathcal{S}^{\pm}_{\mu} = [1 \pm (-1)^{n_\mu}]/2
\ee
for even $(+)$ or odd $(-)$ $n_\mu$ families. For simplicity, in the subsequent discussion, we will focus on even resonance families.  Thus, we consider the interaction:
\be
\label{kernelS}
\mathcal{D}^+_{\sigma \tau}(\mbf{r},\mbf{r}^\prime, z_0) = \Delta_{Q_0}\sum_{\mu}\frac{\Xi_{\mu}(\mbf{r})\Xi_{\mu}(\mbf{r}^\prime)}{\Delta_{\mu}+i\kappa}\mathcal{O}_{\mu}^\sigma \mathcal{O}_{\mu}^{\tau} \mathcal{S}^{+}_{\mu}.
\ee

To evaluate the sums over Gauss-Hermite functions appearing here, we will make repeated use of the Green's function of the harmonic oscillator, which can be written as a sum over Gauss-Hermite functions; namely:
\begin{align}
&G(\vec{r},\vec{r}^\prime,\varphi)
\equiv
\sum_{\mu}\Xi_{\mu}(\vec{r})\Xi_{\mu}(\vec{r}^\prime) e^{-n_\mu \varphi} 
\nonumber\\ 
&\quad= \frac{1}{\pi(1-e^{-2\varphi})}\exp{\left[
-\frac{(\vec{r}^2+{\vec{r}^\prime}^2)/ w_0^2}{\tanh(\varphi)}
+\frac{2\vec{r}\cdot\vec{r}^\prime/w_0^2}{\sinh(\varphi)}
\right]}
\nonumber\\
&\quad= \frac{e^{\varphi}}{2 \pi \sinh(\varphi)}
\exp\left[- \frac{(\vec{r}+\vec{r}^\prime)^2/w_0^2}{2 \tanh(\varphi/2)} -
\frac{(\vec{r}-\vec{r}^\prime)^2/w_0^2}{2 \coth(\varphi/2)}\label{harmonicgreen}
\right].
\end{align}
We can then rewrite the expressions actually required in Eq.~\eqref{kernelS} in terms of this closed-form expression. First, to account for the denominator present in Eq.~\eqref{kernelS}, we define a modified Green's function as:
\begin{align}
\mathcal{G}(\mbf{r},\mbf{r}^\prime,\varphi) &= \displaystyle\sum_{\mu} \frac{\Xi_{\mu}(\mbf{r})\Xi_{\mu}(\mbf{r}^\prime) e^{-n_\mu \varphi}}{1 + \tilde{\epsilon} n_\mu + i \tilde{\kappa}} 
\nonumber\\
&= \int^{\infty}_{0} d \lambda e^{-\lambda (1+i\tilde{\kappa})} 
G(\vec r, \vec r^\prime, \varphi  +\tilde{\epsilon}\lambda),
\label{nongreen}
\end{align}
with $\tilde{\epsilon} = \epsilon/\Delta_{Q_0}$ and $\tilde{\kappa} = \kappa/\Delta_{Q_0}$.  Second, to account for the factor $\mathcal{S}^{\pm}$ in the sum, we can note that $\Xi_{\mu}(-\vec{r}^\prime) = \Xi_{\mu}(\vec{r}^\prime) (-1)^{n_\mu}$, and so we define:
\be
\mathcal{G}^{+}(\mbf{r},\mbf{r}^\prime,\varphi) = \mathcal{G} (\mbf{r},\mbf{r}^\prime,\varphi) + \mathcal{G} (\mbf{r},-\mbf{r}^\prime,\varphi).
\ee
Using these results, we can also include the phase factors arising from the overlaps $\mathcal{O}^\sigma_\mu$ to obtain $\mathcal{D}^+$ in matrix form:
\begin{align}
4 \mathcal{D}^{+} (\mbf{r},\mbf{r}^\prime,z_0) &=  \mathbb{1} \mathcal{G}^{+}(\mbf{r},\mbf{r}^\prime,0) \nonumber \\ &+ \frac{\sigma^{z}}{2} \left[ \mathcal{G}^{+} (\mbf{r},\mbf{r}^\prime,-2 i \theta_0) + \mathcal{G}^{+} (\mbf{r},\mbf{r}^\prime,2 i \theta_0) \right] \nonumber \\ 
&+ \frac{\sigma^{x}}{2i} \left[ \mathcal{G}^{+} (\mbf{r},\mbf{r}^\prime,-2 i \theta_0) - \mathcal{G}^{+} (\mbf{r},\mbf{r}^\prime,2 i \theta_0) \right],
\label{confocalkernel}
\end{align}
where $\mathbb{1},~\sigma^{z}$ and $\sigma^{x}$ are the standard Pauli matrices, and the $z_0$ dependence of this expression comes from the form of $\theta_0$.

As discussed in our previous work~\cite{VaidyaPRX} and explored in the companion paper Ref.~\cite{GouyPRL2018}, $\mathcal{G}^{+}(\mbf{r},\mbf{r}^\prime,0)$ corresponds to the local part of the interaction, while the terms involving $\mathcal{G}^{+}(\mbf{r},\mbf{r}^\prime,\pm 2 i \theta_0)$ gives rise to a sign-changing nonlocal interaction.  On the midplane of the cavity where $\theta_0=\pi/4$, this interaction is proportional to $\cos(2 \vec{r}\cdot\vec{r}^\prime/w_0^2)$.
There are two consequences of the existence of the nonlocal interaction: 1) different phases of the atomic density wave are preferred at different locations in the cavity due to the spatial dependence of $\mathcal{G}^{+} (\mbf{r},\mbf{r}^\prime,2i\theta_0[z_0])$; 2) coupling between $\psi_{c,s}$ is introduced by the nonlocal interaction and atoms may adopt intermediate phases between $\cos(k_rz+\delta)$ and $\sin(k_rz + \delta)$.

The additional freedom in the atomic density wave can be seen directly through the observed light field above the self-organization transition threshold. To show why, we must determine the light field in the cavity and extract the forward propagating part, since we only image the field emitted from one side of the cavity.  From Eq.~\eqref{eq:eom}, we see that within the approximation that allows us to perform adiabatic elimination of cavity modes, the amplitudes of these modes take the form:
\begin{equation}
  \alpha_\mu
  =
  \frac{\Omega g_0 N}{2\sqrt{2}\Delta_a} 
  \int d \mbf{r}^\prime \frac{\Xi_{\mu}(\mbf{r}^\prime)}{\Delta_\mu + i \kappa} \rho(\mbf{r}^\prime)
\displaystyle
\sum_{\sigma=c,s} 
\mathcal{O}_\mu^\sigma \chi_\sigma, 
\label{modeweight}
\end{equation}
where we have defined
\begin{math}
\chi_\sigma \equiv \psi^{}_0\psi^{\ast}_{\sigma} + \psi_0^\ast \psi^{}_{\sigma}
\end{math}.
We can then write the spatially varying light field in the cavity as:
\begin{widetext}
\begin{align}
\tilde{\alpha}(\vec r,z) \equiv& \sum_{\mu} \alpha_{\mu}
\Xi_\mu(\vec{r})  
\cos(k_rz - \Theta_{Q_0}- n_\mu \theta_0 )
= \frac{\Omega g_0 N}{2\sqrt{2}\Delta_a} \int d \mbf{r}^\prime \!\! \sum_{\mu,\sigma=c,s} \!\! \frac{\Xi_{\mu}(\mbf{r})\Xi_{\mu}(\mbf{r}^\prime)}{\Delta_{\mu}+i\kappa} \cos(k_rz - \Theta_{Q_0}- n_\mu \theta_0 )  \mathcal{O}^{\sigma}_{\mu} \chi_\sigma \rho(\mbf{r}^\prime) \nonumber \\
\propto& \frac{1}{2} e^{i(k_rz - \Theta_{Q_0})} \int d \mbf{r}^\prime \left\{ \chi_c \left[ \mathcal{G}^{+}(\mbf{r},\mbf{r}^\prime,0) + \mathcal{G}^{+}(\mbf{r},\mbf{r}^\prime,-2 i \theta_0) \right] + i \chi_s \left[ \mathcal{G}^{+}(\mbf{r},\mbf{r}^\prime,0) - \mathcal{G}^{+}(\mbf{r},\mbf{r}^\prime,-2 i \theta_0) \right] \right\} \rho(\mbf{r}^\prime) + \nonumber \\
& \frac{1}{2} e^{-i(k_rz - \Theta_{Q_0})} \int d \mbf{r}^\prime \left\{ \chi_c \left[ \mathcal{G}^{+}(\mbf{r},\mbf{r}^\prime,0) + \mathcal{G}^{+}(\mbf{r},\mbf{r}^\prime,2 i \theta_0) \right] - i \chi_s \left[ \mathcal{G}^{+}(\mbf{r},\mbf{r}^\prime,0) - \mathcal{G}^{+}(\mbf{r},\mbf{r}^\prime,2 i \theta_0) \right] \right\} \rho(\mbf{r}^\prime), \nonumber \\
\end{align}
\end{widetext}
where again we have used Eq.~\eqref{nongreen} to rewrite the sum over transverse modes. The forward traveling component of the light field can then be rewritten as
\begin{align}
\tilde{\alpha}^F(\vec{r}) \propto& \int d \mbf{r}^\prime \rho(\mbf{r}^\prime) \mathcal{G}^{+}(\mbf{r},\mbf{r}^\prime,0)  \nonumber \\
&+e^{-i2\phi_A}\int d \mbf{r}^\prime \rho(\mbf{r}^\prime) \mathcal{G}^{+}(\mbf{r},\mbf{r}^\prime,-2 i \theta_0),
\label{forwardlight}
\end{align}
where $\phi_A = \mathrm{Arg}[\chi_c + i \chi_s]$.  Thus, while the light \emph{inside} the cavity is purely real, the forward travelling wave, and thus the cavity light emitted out of one side of the cavity, contains important phase information.
Physically, $\phi_A$ corresponds to the phase of the density wave adopted by the atoms. In the cavity output field, $\mathcal{G}^{+}(\mbf{r},\mbf{r}^\prime,0)$ and $\mathcal{G}^{+}(\mbf{r},\mbf{r}^\prime,-2i\theta_0)$ correspond to two distinct spatial features: the former results in an intense localized spot at the position of the atoms, while the latter gives rise to an weak oscillating background $\cos(2\vec{r}\cdot\vec{r}^\prime/w_0^2)$, as shown below. Both have been observed~\cite{VaidyaPRX,GouyPRL2018}. Therefore, measuring the phase difference between these two parts of imaged light field,  $\Delta_\phi = - 2 \phi_A$,  reveals the phase offset of the atomic density wave relative to the cavity modes, as explored in the companion paper Ref.~\cite{GouyPRL2018}.

\section{$\mathbb{U}(1)$-symmetry in an ideal confocal cavity}\label{U1sym}
We now consider the case of an ideal confocal cavity (i.e., all modes are perfectly degenerate $\epsilon=0$) and show the emergence of a $\mathbb{U}(1)$-symmetry for the phase of the density waves in the self-organization transition. Moreover, we will consider the case where there is no cavity loss ($\kappa = 0$).  Together, these  allow the  replacement of $\mathcal{G}^{+}$ in the previous section with the symmetrized, bare harmonic oscillator Green's function
\be
G^{+}(\mbf{r},\mbf{r}^\prime,\varphi) = G(\mbf{r},\mbf{r}^\prime,\varphi) + G(\mbf{r},-\mbf{r}^\prime,\varphi)
\ee
with $G(\vec{r},\vec{r}^\prime,\varphi)$ as defined in Eq.~\eqref{harmonicgreen}.
We note that for atoms trapped at the midplane of the cavity, $z_0=0$ so $\theta_0=\pi/4$, and so the nonlocal interaction now reduces to $G^{+} (\mbf{r},\mbf{r}^\prime,i \pi/2)$ and can be directly evaluated using Eq.~\eqref{harmonicgreen} as
\be
G^{+} (\mbf{r},\mbf{r}^\prime,i \pi/2) = \frac{1}{\pi}\cos \left( \frac{2 \mbf{r} \cdot \mbf{r}^\prime}{w^2_0}\right),
\ee
which is the nonlocal interaction studied in the companion paper~\cite{GouyPRL2018}.  This expression is purely real, and so by inserting this into Eq.~\eqref{confocalkernel}, we see that
there is no $\sigma^x$ term.  Moreover, at $2r^2/w_0^2=(n+1/2)\pi$, with integer $n$, the self-interaction $4 \Re \{\mathcal{D}^{+}(\vec r, \vec r)\} = \mathbb{1} G^{+}(\mbf{r},\mbf{r},0)$ will become proportional to the identity matrix, indicating complete freedom about what the relative amplitudes of the sine and cosine components of the density wave can be.  That is, these two density wave patterns become degenerate.

\begin{figure}
    \centering
    \includegraphics[width = 0.49\textwidth]{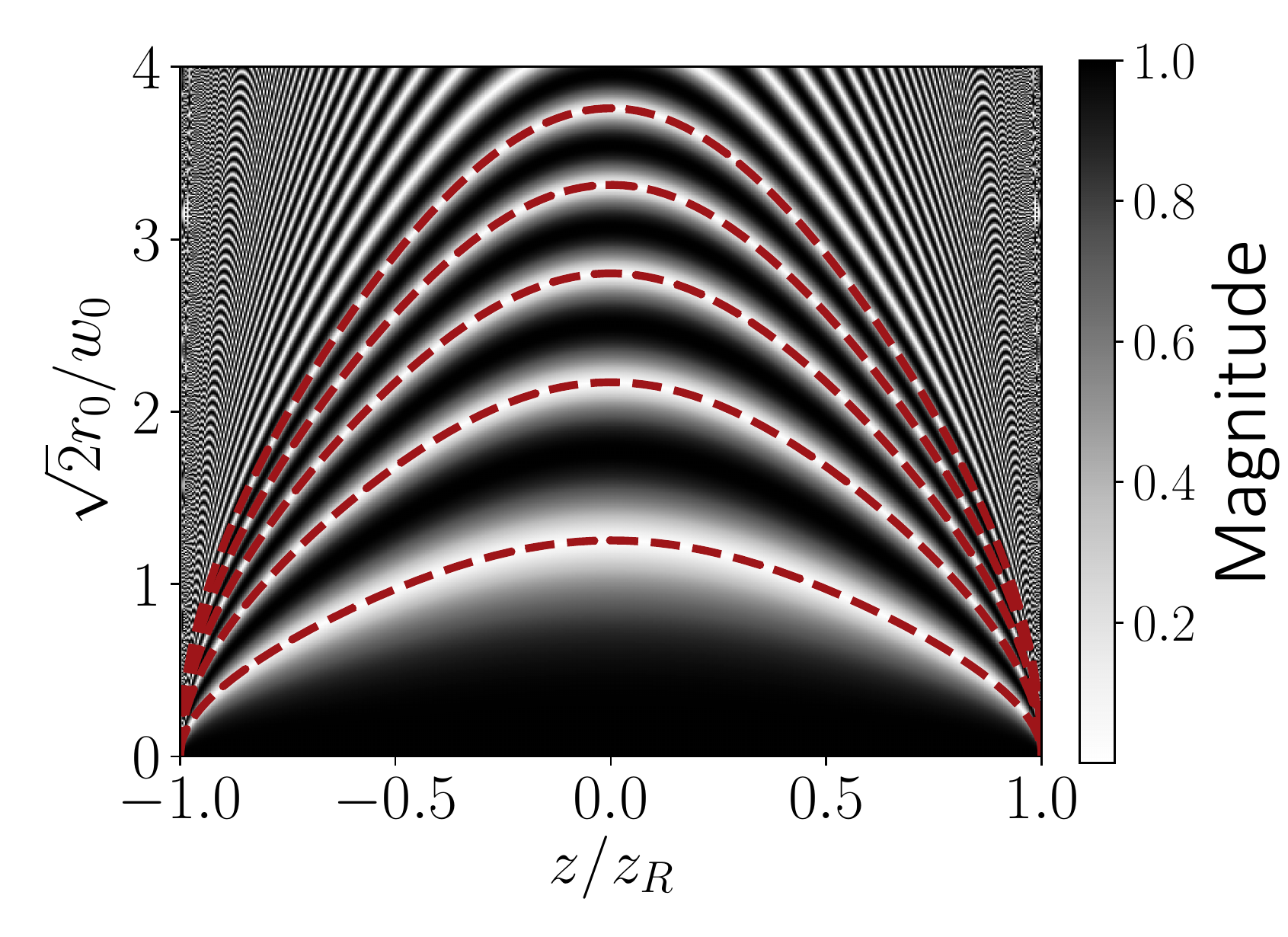}
    \caption{Magnitude of  $|G^+(\mbf{r}_0,\mbf{r}_0,2 i \theta_0(z_0)) |$ as a function of radius  $r_0$ (vertical axis) and distance along the cavity  $z$ (horizontal axis).  Red dashed lines mark solutions to Eq.~\eqref{degenerateradii}, indicating contours at which there is a degeneracy between the sine and cosine density wave patterns.}
    \label{fig:sketchmodG}
\end{figure}

If we return to consider a general value $\theta_0$, and thus a general longitudinal location $z_0$ of the atomic gas, we can explore how these zeros of the non-local interaction matrix evolve. We consider here the case of a single atomic cloud where the Thomas-Fermi radius of the cloud is much smaller than the cavity waist $w_0$ and the density profile can be treated as a $\delta$-function. For a cloud centered at location $z_0$ and $\mbf{r}_0$, with density profile $\delta(\mbf{r}-\mbf{r}_0)$ in the cavity transverse plane, the interaction matrix given by
\begin{multline}\label{intmatrixreim}
4\Re \{\mathcal{D}^{+}(\mbf{r}_0,\mbf{r}_0,z_0) \} = \mathbb{1} G^{+}(\mbf{r}_0,\mbf{r}_0,0) 
+\\
\sigma^z \Re \{G^{+}(\mbf{r}_0,\mbf{r}_0,-2 i \theta_0) \}  
+
\sigma^x \Im \{G^{+}(\mbf{r}_0,\mbf{r}_0,-2 i \theta_0) \}.
\end{multline}
Note here, we have used that $G^+(\mbf{r}_0,\mbf{r}_0,-2 i \theta_0)^\ast =G^+(\mbf{r}_0,\mbf{r}_0,2 i \theta_0)$, a result that holds for the bare Green's function $G^{+}(\mbf{r}_0,\mbf{r}_0,\varphi)$ but does not directly hold for the general form $\mathcal{G}(\mbf{r}_0,\mbf{r}_0,\varphi)$; we return to this point below.
As such, we note that a degeneracy between $\psi_c$ and $\psi_s$ occurs when
\be
|G^+(\mbf{r}_0,\mbf{r}_0,2 i \theta_0) | = 0.
\label{magicradius}
\ee
Neglecting all prefactors, we find this condition equivalent to:
\be
 \Bigl|
  e^{- i 2 r_0^2 \tan(\theta_0)/w_0^2 }
  +
  e^{+ i 2 r_0^2 \cot(\theta_0)/w_0^2 }
  \Bigr| = 0.
\ee
Using the fact that
\be
  \tan[\theta_0(z_0)]=
  \tan\left[ \frac{\pi}{4} + \text{arctan}\left( \frac{z_0}{z_R} \right) \right]
  = \frac{z_R + z_0}{z_R - z_0},
\ee
we can directly find the equation for radii at which degeneracy between the sine and cosine density wave patterns occurs,
\be
\sqrt{2}r_0/w_0  = 
  \sqrt{\left[
    \frac{z_R^2 - z_0^2}{z_R^2 + z_0^2}
  \right]
  \pi
  \left(
    n +
    1/2
  \right)}, 
  \label{degenerateradii}
\ee
where $n$ is a non-negative integer that indexes the family of possible radii.  We readily see that for $z_0=0$ this reproduces the zeros of $\cos(2 r^2/w_0^2)$. The  contours for other positions $z_0$  are illustrated in Fig.~\ref{fig:sketchmodG}.

To understand the consequence of such degeneracy, it is useful to consider the effective Hamiltonian corresponding to Eq.~\eqref{eq:atomeom}. Neglecting the bare atomic Hamiltonian, at a radius $r_0$ satisfying Eq.~\eqref{degenerateradii}, the effective Hamiltonian describing the cavity-mediated interaction is given by
\be
H_{\mrm{eff}} \propto \begin{pmatrix} \chi_c & \chi_s \end{pmatrix} \begin{pmatrix} G^{+}(\mbf{r}_0,\mbf{r}_0,0) & 0 \\
0 & G^{+}(\mbf{r}_0,\mbf{r}_0,0)\end{pmatrix}
\begin{pmatrix}
\chi_c \\
\chi_s
\end{pmatrix}.
\ee
We can now see that the Hamiltonian is invariant under the transformation
\be
\begin{pmatrix}
\psi_c \\
\psi_s
\end{pmatrix}
\rightarrow
\mathbf{R}(\phi)
\begin{pmatrix}
\psi_c \\
\psi_s.
\end{pmatrix}
, \quad
\mathbf{R}(\phi)
\equiv
\begin{pmatrix}
\cos \phi & -\sin \phi \\
\sin \phi & \cos \phi
\end{pmatrix}.
\ee
Physically, the transformation corresponds to the freedom in the phase of the atomic density wave, i.e., any density wave $\cos(k z  + \delta)$ with arbitrary $\delta$ is allowed.  This degeneracy is analogous to the situation in a recent experimental realization of a supersolid via the  coupling of a BEC to two crossed cavities~\cite{Leonard2017}. The spontaneous breaking of this $\mathbb{U}(1)$-symmetry can be directly observed by imaging the phase of the forward-traveling component of the cavity light field, as can be seen from Eq.~\eqref{forwardlight}. 

So far we have only showed that the symmetry exists when we ignore the contribution quadratic in the light field $\Xi_{\mu}(\vec r)$ in the equation of motion. As such, one may question whether the symmetry should survive far above threshold, where there is scattering between different cavity modes (in the second term on the right-hand side of Eq.~\eqref{eq:eom}). As we will next show, the symmetry in fact exists even when such coupling terms are included, as long as we remain in the confocal limit.  To see this, we  rewrite the initial atom-cavity Hamiltonian  in terms of new photon field operators.  Specifically, we define the new operators
\begin{displaymath}
  \hat{a}_{c,s}
  =\frac{1}{\sqrt{\mathcal{N}_{c,s}}} \sum_\mu  \hat{a}_\mu^{} \Xi_\mu(\vec r_0) \mathcal{S}^{+}_\mu
  \begin{cases}
    \cos(n_\mu \theta_0)\\
    \sin(n_\mu \theta_0)
  \end{cases},
\end{displaymath}
where $\mbf{r}_0$ is the position of the atomic cloud. The normalization factor $\mathcal{N}_{c,s}$ is required to impose bosonic commutation relations on the cavity modes. Computing the commutation relations, we obtain
\begin{align}
  [ \hat{a}_c, \hat{a}^\dagger_s] &=\frac{1}{2\sqrt{\mathcal{N}_c \mathcal{N}_s}} \Im \{ G^{+}(\vec r_0, \vec r_0,   2i \theta_0) \}  \\
  [ \hat{a}_c, \hat{a}^\dagger_c] 
  &=
  \frac{1}{2 \mathcal{N}_c^2 } \left[
    G^{+}(\vec r_0, \vec r_0, 0)
    +
    \Re \{G^{+}(\vec r_0, \vec r_0,  2 i\theta_0) \}
  \right]
  \nonumber \\
  [ \hat{a}_s, \hat{a}^\dagger_s]
  &=
  \frac{1}{2 \mathcal{N}_s^2} \left[
    G^{+}(\vec r_0, \vec r_0, 0)
    -
   \Re \{ G^{+}(\vec r_0, \vec r_0,   2i \theta_0) \}\nonumber
  \right].
\end{align}
We may now observe that the cross-commutator necessarily vanishes when the location $\mbf{r}_0$ of the atoms satisfies the condition in Eq.~\eqref{magicradius},  $|G^+(\vec r_0, \vec r_0, i 2 \theta_0)| = 0$.  This implies that the  normalizations required to satisfy the other two commutation relations are equal, $\mathcal{N}_s = \mathcal{N}_c$.  We will see that this equality leads to the emergence of the $\mathbb{U}(1)$-symmetry of interest.

Using the above definitions, along with the parameterization in Eq.~\eqref{atomwf}, we can rewrite the matter-light coupling terms in the original Hamiltonian Eq.~\eqref{eq:Hamiltonian} as 
\begin{widetext}
\begin{multline}
H_{\mrm{coupling}} = \frac{g_0N\Omega}{2\sqrt{2}\Delta_a}
\displaystyle\sum_{\sigma=c,s}
\sqrt{\mathcal{N}_\sigma}\psi_0(\psi^{\ast}_{\sigma}+\psi_\sigma^{})
(\hat{a}^{\dagger}_\sigma+\hat{a}^{}_\sigma) 
+ \frac{g^2_0 N}{2 \Delta_a} 
\Big[ 
\mathcal{N}_c \hat{a}^{\dagger}_c \hat{a}_c 
\left(\psi^2_0 + \frac{3}{4} |\psi_c|^2+ \frac{1}{4} |\psi_s|^2\right) 
\\+ \mathcal{N}_s \hat{a}^{\dagger}_s \hat{a}_s
\left(\psi^2_0 + \frac{3}{4} |\psi_s|^2+ \frac{1}{4} |\psi_c|^2\right) + 
\frac{1}{4}\sqrt{\mathcal{N}_c \mathcal{N}_s}
(\psi^{\ast}_c \psi_s + \psi^{\ast}_s \psi_c)
(\hat{a}^{\dagger}_c \hat{a}_s + \hat{a}^{\dagger}_s \hat{a}_c) \Big].
\end{multline}
\end{widetext}
The various factors of $1/4$ and $3/4$ come from the terms $\langle \cos^4(k_rz)\rangle=\langle\sin^4(k_rz)\rangle = 3/8$ and $\langle \cos^2(k_rz) \sin^2(k_rz) \rangle = 1/8$ encountered when taking the averages of the Stark shifts.
When $\mathcal{N}_s = \mathcal{N}_c=\mathcal{N}$, we can rewrite the term in brackets as
\begin{multline*}
\mathcal{N}\Big[ 
\left( \hat{a}^{\dagger}_c \hat{a}_c  + \hat{a}^{\dagger}_s \hat{a}_s\right)
\left(\psi^2_0 + \frac{1}{2} |\psi_c|^2+ \frac{1}{2} |\psi_s|^2\right) 
\\+
\frac{1}{4}
\left( \hat{a}^{\dagger}_c \hat{a}_c  - \hat{a}^{\dagger}_s \hat{a}_s\right)
\left(|\psi_c|^2 - |\psi_s|^2\right) 
\\+
\frac{1}{4}
(\psi^{\ast}_c \psi_s + \psi^{\ast}_s \psi_c)
(\hat{a}^{\dagger}_c \hat{a}_s + \hat{a}^{\dagger}_s \hat{a}_c) \Big]
\end{multline*}
and then readily verify that the Hamiltonian is invariant under the combined rotation
\be
\begin{pmatrix}
a_c \\
a_s 
\end{pmatrix} \rightarrow 
\mathbf{R}(\phi)
\begin{pmatrix}
a_c \\
a_s 
\end{pmatrix}, 
\quad
\begin{pmatrix}
\psi_c \\
\psi_s 
\end{pmatrix} \rightarrow 
\mathbf{R}(\phi)
\begin{pmatrix}
\psi_c \\
\psi_s 
\end{pmatrix}, 
\ee
Hence,  it appears that the $\mathbb{U}(1)$-symmetry remains robust in a perfect confocal cavity even when including the nonlinear terms that are relevant far above threshold.

\section{Breaking of emergent\\$\mathbb{U}(1)$-symmetry}
\label{breaking-u1}

The previous section assumes a perfect confocal cavity, with no loss and no transverse mode dispersion. In that case, we anticipated perfect $\mathbb{U}(1)$-symmetry at special radii.  Experimentally, this should correspond to a shot-to-shot fluctuating density wave phase for these radii.  As discussed below, this is not seen experimentally.  Instead, there is a window where the phase of the density wave evolves smoothly and deterministically from $\phi_A=0$ to $\phi_A=\pi/2$.  To understand this, we next discuss the effects beyond the ideal solution, specifically effects of a finite cloud size, effects of displacement from the midpoint of the cavity, and effects of a small mode-splitting $\epsilon$ and cavity loss $\kappa$.

To study the emergence of symmetry-breaking terms, we consider the scenario where a single BEC is placed at a radius $\mbf{r}_0$ in the cavity transverse plane, and we determine the phase of density wave formed in the self-organization transition from the eigenvectors of the effective interaction matrix.  The effective interaction matrix is the generalization of Eq.~\eqref{confocalkernel} to include the aforementioned deviations from an ideal confocal cavity. 

\subsection{Displacement from midplane of cavity $z_0=0$}

Displacing the BEC from the midpoint  $z_0=0$ of the cavity induces a density wave that is, in general, neither of purely sine nor cosine pattern except at special positions where complete phase freedom ($\mathbb{U}(1)$-symmetry) is restored.   Admixtures of sine and cosine patterns are favored at positions where  $G^+(\vec{r}_0,\vec{r}_0, 2 i \theta_0)$ is complex. This function is complex  at all points in the cavity except at its midpoint  where  $G^+(\vec{r}_0,\vec{r}_0, i \pi/2)$ is purely real, as well as at the points indicated by the red dashed line contours in Fig.~\ref{fig:sketchmodG} where  $G^+(\vec{r_0},\vec{r}_0, 2 i \theta_0)=0$, resulting in positions of restored phase freedom.  As such, displacement from the midplane of the cavity does not break the $\mathbb{U}(1)$-symmetry at these specific radii.

In the following, we will see that finite cloud size and nonconfocality can alter this picture. These nonidealities lift the pattern degeneracy even at these special positions. What is left is instead  a smooth, deterministic evolution of density wave phase with no $\mathbb{U}(1)$-symmetry points.  The exception to this is at midplane $z_0=0$. That is, despite these nonidealities, the interaction remains purely real  at the midpoint of the cavity because $\theta_0 =\pi/4$ at  $z_0=0$.  Therefore, at the midplane of the cavity, there continue to exist special radii where the $\mathbb{U}(1)$-symmetry is expected even under conditions of finite cloud size and nonconfocality.

We shall focus on the nonlocal contribution $\mathcal{G}^{+} (\mbf{r}_0,\mbf{r}_0,2 i \theta_0)$  to the self-interaction, since  the effects of nonconfocality on the local interaction $\mathcal{G}^{+} (\mbf{r}_0,\mbf{r}_0,0)$ have been studied in detail in Ref.~\cite{VaidyaPRX}. Moreover, as seen in Eq.~\eqref{confocalkernel}, the density wave  phase is determined not by the local terms, which scale the components of the identity matrix, but by the nonlocal terms that  determine the $\sigma^x$ and $\sigma^z$ parts of the interaction matrix.

\subsection{Nonconfocality}

To consider nonconfocality, we return to using the Green's functions defined in Eq.~\eqref{nongreen}, 
i.e., allowing for nonzero $\epsilon$ and $\kappa$.
We first note that by using this Green's function, the matrix form of the nonlocal cavity-mediated interaction Eq.~\eqref{confocalkernel} can be put into the form 
\begin{multline}
4 \mathcal{D}^{+}_{\text{nonlocal}}(\mbf{r}_0,z_0) =  \\
\int^{\infty}_0 \!d \lambda e^{-\lambda(1 + i \tilde\kappa)}
 \Big[
 \sigma^z \Re\{ G^{+}(\mbf{r}_0,\mbf{r}_0,\varphi - 2 i\theta_0)\}
 \\
+\sigma^x \Im \{G^{+}(\mbf{r}_0,\mbf{r}_0,\varphi - 2 i\theta_0)\}
\Big],
\label{eq:nonconfocalcomplexD}
\end{multline}
where once again we note the $z_0$ dependence comes from the form of $\theta_0$.
We further note that away from $\theta_0(0)=\pi/4$, the quantity 
$G^{+} (\mbf{r}_0,\mbf{r}_0,-2i\theta_0)$ is complex, so both components exist (except at the special points mentioned above where it is zero).  We can simplify this further, since the actual interaction appearing in Eq.~\eqref{eq:atomeom} involves $\Re\{\mathcal{D}^+\}$. Since the term in brackets in Eq.~\eqref{eq:nonconfocalcomplexD} is by definition real, taking the real part thus replaces
$e^{-i \lambda \tilde{\kappa}} \to \cos(\tilde{\kappa}\lambda)$.  We can then write the final form for the relevant part of the interaction as:
\begin{align}
4 \Re \{ \mathcal{D}_{\mrm{nonlocal}}^{+} (\mbf{r}_0,z_0) \}&= 
\sigma^z  \Re\{\widetilde{G}^{+}(\mbf{r}_0,\mbf{r}_0, - 2 i \theta_0)\}
\nonumber\\&+
\sigma^x  \Im\{\widetilde{G}^{+}(\mbf{r}_0,\mbf{r}_0, - 2 i \theta_0)\}
\label{nonmatrix}
\end{align}
where we have defined a nonconfocal Green's function:
\begin{equation}
\widetilde{G}^{+}(\mbf{r},\mbf{r}^\prime, \varphi)
=
\int^{\infty}_0 \!\!d \lambda e^{-\lambda} \cos(\tilde{\kappa}\lambda) 
  G^{+}(\mbf{r},\mbf{r}^\prime,\tilde\epsilon \lambda +\varphi).
\label{tildeG}
\end{equation}

From this expression, we can see that  we still have an interaction that is purely of $\sigma^z$ form at $\theta_0=\pi/4$.  To see this, we note that 
\begin{multline*}
\Im\{ G^{+}(\mbf{r},\mbf{r}^\prime,\tilde\epsilon \lambda - i \pi/2)\} \\\equiv
\Im\left\{\sum_{\mu}\Xi_{\mu}(\vec{r})\Xi_{\mu}(\vec{r}^\prime) \mathcal{S}_\mu^+ e^{-n_\mu (\varphi - i \pi/2)} \right\}
\\=
\sum_{\mu}\Xi_{\mu}(\vec{r})\Xi_{\mu}(\vec{r}^\prime) \mathcal{S}_\mu^+ e^{-n_\mu \varphi} \sin( n_\mu \pi/2).
\end{multline*}
Because $\mathcal{S}_\mu^+$ restricts the sum to even terms, we immediately see that $\sin(n_\mu \pi/2)=0$ for these terms so all terms vanish. The vanishing imaginary part then means the interaction matrix becomes purely diagonal and there still exists points where the interaction is proportional to the identity matrix, giving degeneracy and the $\mathbb{U}(1)$-symmetry. However, away from $\theta_0=\pi/4$, one can readily check that $|\widetilde{G}^+(\vec{r_0},\vec{r}_0, 2 i \theta_0)|$ does not  vanish, i.e., the real and imaginary parts of this function no longer vanish at the same points.  As a result, there would be a deterministic phase at all radii (other than those at the midpoint), with the sense of phase winding dependent on the displacement from the midpoint of the cavity.

\subsection{Finite transverse size of atomic gas}

A second effect that also leads to a deterministic phase winding is the finite size of the condensate.  To capture this, we consider averaging the interaction matrix over the spatial profile of the cloud.  That is, we define
an averaged function
\be
G^+_{\text{avg}}(\mbf{r}_0,\varphi) \equiv
\int\!d\mbf{r} \int\!d\mbf{r}^\prime \rho(\mbf{r}^\prime) G^{+}(\mbf{r},\mbf{r}^\prime,\varphi) \rho(\mbf{r}),
\label{eq:selfG}
\ee
and then define the corresponding interaction matrix, describing the self-interaction due to the non-local part of the interaction,
\begin{align}
4 \Re \{ \mathcal{D}_{\mrm{avg,nonlocal}}^{+} (\mbf{r}_0,z_0) \}&= 
\sigma^z  \Re\{{G}^{+}_{\text{avg}}(\mbf{r}_0,\mbf{r}_0, - 2 i \theta_0)\}
\nonumber\\&+
\sigma^x  \Im\{{G}^{+}_{\text{avg}}(\mbf{r}_0,\mbf{r}_0, - 2 i \theta_0)\},
\label{selfmatrix}
\end{align}
where once again the $z_0$ dependence comes from the form of $\theta_0$.

The integral over coordinates $\vec r, \vec r^\prime$ can be evaluated by using a Gaussian density profile:
\be
\rho(\mbf{r}-\mbf{r}_0) = \frac{1}{2 \pi \sigma_A^2} \exp\left(\frac{-(\mbf{r} - \mbf{r}_0)^2}{2 \sigma_A^2}\right),
\ee
where $\sigma_A$ is the BEC width, and by making use of the harmonic oscillator Green's function Eq.~\eqref{harmonicgreen}.  We find the integral over radius yields
\begin{multline}
G^+_{\text{avg}}(\mbf{r}_0,\varphi) = 
\frac{A/\pi}{1-e^{-2 \varphi}}
\Big\{
\mrm{exp} \left[-\frac{2 \mbf{r}^2_0}{2 \sigma_A^2 +  \coth(\varphi/2) w^2_0  } \right] \nonumber \\+ 
\mrm{exp} \left[-\frac{2 \mbf{r}^2_0}{2 \sigma_A^2 +  \tanh(\varphi/2) w^2_0  } \right] \Big\},
\end{multline}
where the prefactor $A$ takes the form
\begin{equation}
    A = \left[
    \left(1+\frac{2 \sigma_A^2}{w_0^2 \coth(\varphi/2)}\right)
    \left(1+\frac{2 \sigma_A^2}{w_0^2 \tanh(\varphi/2)}\right)
    \right]^{-1}.
\end{equation}

\begin{figure}
    \centering
    \includegraphics[width = 0.49\textwidth]{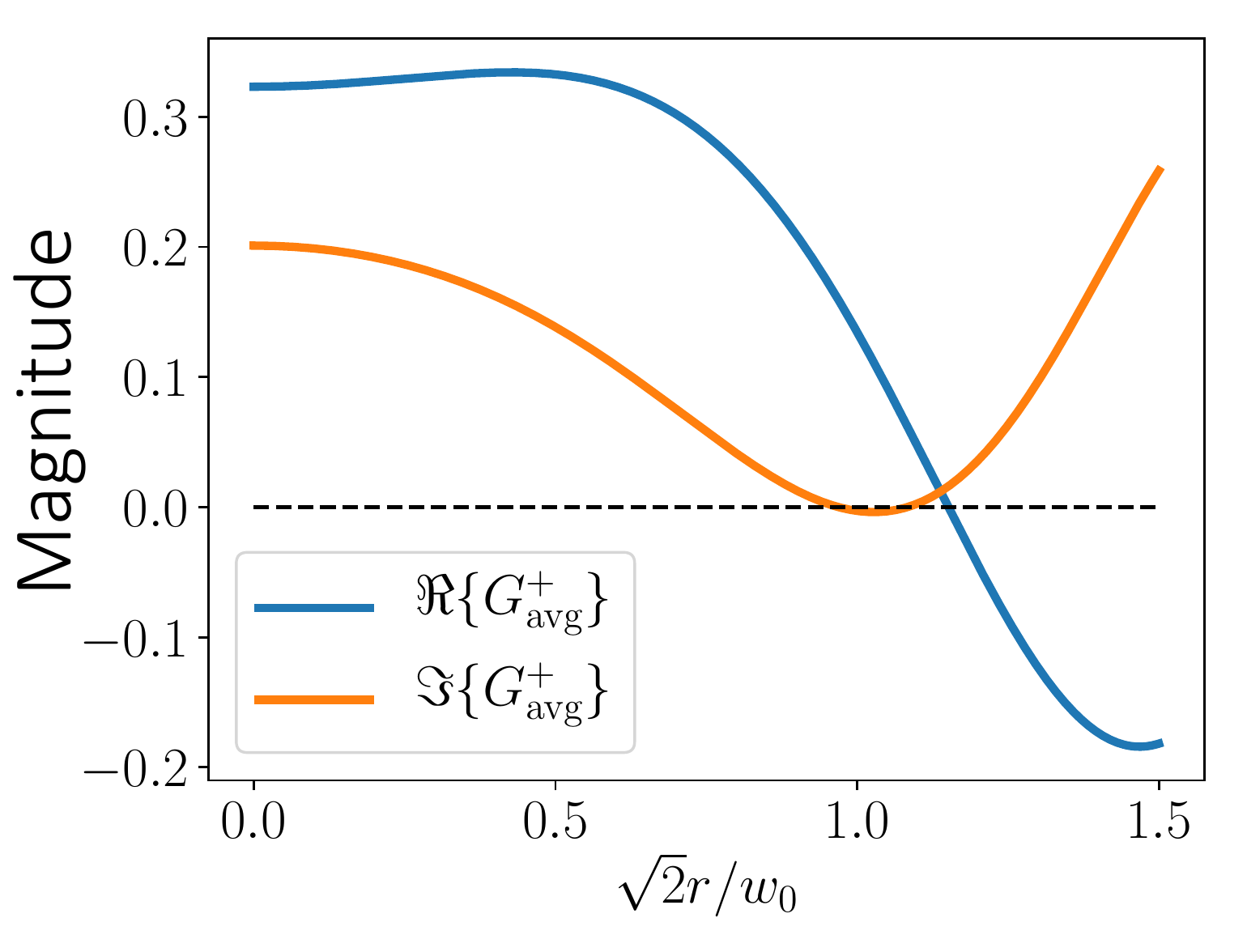}
    \caption{Real and imaginary parts of $G^{+}_{\mrm{avg}}$ for a BEC  at  a transverse plane ($z_0/z_R = 0.3$) located away from the cavity midplane and with a Gaussian width of $\sigma_A/w_0 = 0.1$. Black dashed line marks zero. Note that $\Re \{ G^{+}_{\mrm{avg}}  \}$ and $\Im \{ G^{+}_{\mrm{avg}}  \}$ do not vanish at the same point.}
    \label{Gplus_finite}
\end{figure}

One can then combine the effects of nonconfocality with finite cloud size by using $\widetilde{G}(\vec r, \vec r^\prime, \varphi)$ in the right-hand side of Eq.~\eqref{eq:selfG}. In practice, one first performs the Gaussian integral defined above and then defines
\begin{equation}
\widetilde{G}^{+}_{\mrm{avg}}(\mbf{r}_0, \varphi)
=
\int^{\infty}_0 \!\!d \lambda e^{-\lambda} \cos(\tilde{\kappa}\lambda) 
  G^{+}_{\text{avg}}(\mbf{r}_0,\tilde\epsilon \lambda +\varphi),
\label{tildeGself}
\end{equation}
which can be inserted into Eq.~\eqref{selfmatrix} to find the effective interaction. The final integral over $\lambda$ is  performed numerically. We note that even when $\tilde{\epsilon} = 0$, the finite extent $\sigma_A$ of the atoms already has the effect of turning $G^{+}_{\mrm{avg}} (\vec{r}_0, -2 i \theta_0)$ into a complex quantity  and ensuring that $|G^{+}_{\mrm{avg}} (\vec{r}_0, -2 i \theta_0)|$ does not vanish
when away from the cavity midplane. Figure~\ref{Gplus_finite} plots the real and imaginary part of $G^{+}_{\mrm{avg}} (-2 i \theta_0)$ for atoms located at $z_0/z_R = 0.3$. Crucially, the real part and imaginary part do not vanish at the same point, therefore there is always a preferred phase of the atomic density wave and, consequently, no emergent $\mathbb{U}(1)$-symmetry. 

\begin{figure}[t!]
\includegraphics[width = 0.49\textwidth]{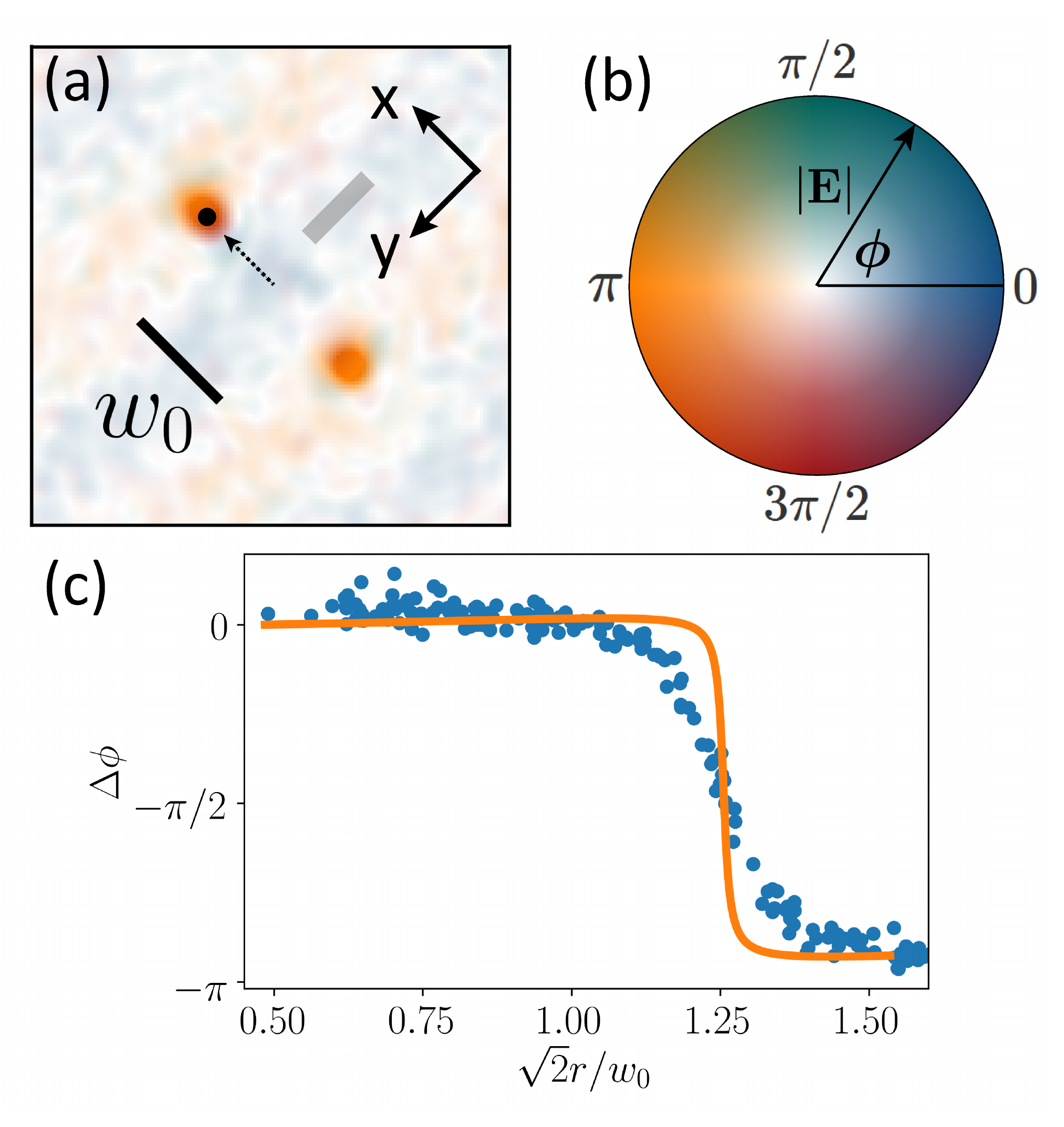}
\caption{(a) Example of the experimental phase and amplitude of the cavity emission.  The waist $w_0$ of the TEM$_{00}$ in the image plane is indicated.   $\Delta \phi=-2\phi_A$ is computed from the difference in  phase between the regions marked by black dot and the average phase of the gray rectangular region. They correspond to the local $\mathcal{G}^{+}(\mbf{r},\mbf{r}^\prime,0)$ and nonlocal $\mathcal{G}^{+}(\mbf{r},\mbf{r}^\prime,2 i \theta_0)$ contribution to the light field in Eq.~\eqref{forwardlight}, resp. (b) Color wheel for illustrating electric field amplitude and phase. (c) Measured phase difference between the local and nonlocal contribution to the cavity field as a function of the distance $r$ of a single BEC from the cavity axis. The data shown here are taken with cavity detuning $\Delta_{Q_0} = -120~\mrm{MHz}$. The distance is extracted by fitting to the amplitude of the measured field. The path of $\vec{r}$ taken in the $x$-$y$ plane from the cavity axis is shown as an black dashed arrow in panel (a).}
\label{winding}
\end{figure}

\subsection{Comparison to experiment}

To confirm the existence of $\mathbb{U}(1)$-symmetry-breaking, we perform an experiment similar in method to those reported in our works reported in Refs.~\cite{VaidyaPRX,GouyPRL2018}. In brief, we create a BEC of $^{87}$Rb with population $2.5(3) {\times} 10^5$ inside the cavity; see Ref.~\cite{Kollar2015} for technical details.  The BEC is confined in an optical tweezer trap of trap frequencies $(\omega_x,\omega_y,\omega_z) = 2 \pi \times [189(2),134(1),90(1)]$~Hz and Thomas-Fermi radii $(R_x, R_y, R_z) = [4.2(1), 5.8(3), 8.9(1)]$~$\mu$m. Due to experimental constraints~\footnote{Geometrical constraints imposed by the vacuum chamber prevented the trapping of the BEC at the exact cavity midplane.}, the BEC is fixed at a position of $z_0 \approx 240~\mu$m from the cavity midplane.  The tweezer allows us to control $\vec{r}_0$ by moving the BEC in the transverse plane.  The confocal cavity has length $L=1$~cm and $\mathrm{TEM}_{00}$ mode waist $w_0=35$~$\mu$m.  For this system, $g_0 = 2 \pi \times 1.47(3)$~MHz and $\kappa = 2 \pi \times 167(4)$~kHz.

To measure the $\mathbb{U}(1)$-symmetry-breaking, we observe the phase difference $\Delta \phi$ between the cavity emission arising from the local versus nonlocal interaction terms while moving a single BEC across a node of the oscillatory cosine pattern due to $\mathcal{G}^{+}(\mbf{r},\mbf{r}',2i\theta_0)$. The amplitude and phase  of the cavity field emission are detected through the holographic reconstruction of a spatial heterodyne measurement; see Refs.~\cite{kroeze2018spinor,GouyPRL2018} for details.  The detection  of the density wave phase relies on the fact that the phase difference $\Delta \phi$  is related to the density wave pattern through $\Delta \phi=-2\phi_A$ using Eq.~\eqref{forwardlight}. That is, the observable $\Delta \phi$ is directly related to the phase of the density wave. 

Figure~\ref{winding} shows the results of such measurements.  An example of a  holographically reconstructed cavity emission field is shown in Fig.~\ref{winding}(a).  The two orange spots are the emission due to the local interaction term  $\mathcal{G}^{+}(\mbf{r},\mbf{r}^\prime,0)$ centered about the real and mirror images of the BEC in the object plane, while the orange and blue striped oscillation arises from the nonlocal term $\mathcal{G}^{+}(\mbf{r},\mbf{r}^\prime,2 i \theta_0)$.

Figure~\ref{winding}(c) shows how $\Delta \phi$ evolves versus $r$.  A particular direction of the phase winding is preferred as the BEC is gradually moved away from the cavity midplane.
As discussed in the beginning of this section, rather than observe a point of random shot-to-shot density wave phase due to a $\mathbb{U}(1)$-symmetry, we observe that the phase smoothly winds from $0$ to $-\pi$ as the BEC crosses the node at the point $\sqrt{2} r/w_0 \approx \sqrt{\pi/2} \approx 1.25$, as expected from Eq.~\eqref{degenerateradii}. The orange solid line is the predicted theoretical phase difference taking into account: 1) the finite mode dispersion $\tilde{\epsilon}$; 2) finite size $\sigma_A$ of the cloud in the transverse plane; and 3) the aforementioned displacement $z_0$ of the atoms in $\hat{z}$ from the cavity midplane.

Our theory could not reproduce the width of the transition region in the density wave phase. We attribute this discrepancy to the coupling between density waves through the nonlinear atom-cavity coupling term that is proportional to $g^2_0/\Delta_a$ in Eq.~\eqref{eq:eom}, which is not taken into account at the current level of our theory. However, our experimental data is taken above the self-organization transition threshold with macroscopic population of cavity photons. In this regime, scattering between different cavity modes is no longer negligible. We also note that we have assumed paraxial optics and only taken into account of the linear dispersion between transverse modes. Contributions from cavity mirror aberrations (e.g., spherical aberration) could further complicate the phase dependence of transverse modes beyond the included Gouy phase shifts.  Lastly, we note that the finite extent of the BEC along the cavity axis could play a role.  We leave a discussion of this effect to the Appendix.

\section{Deterministic evolution of the density wave phase}\label{extensions}

\begin{figure}
\includegraphics[width = 0.49\textwidth]{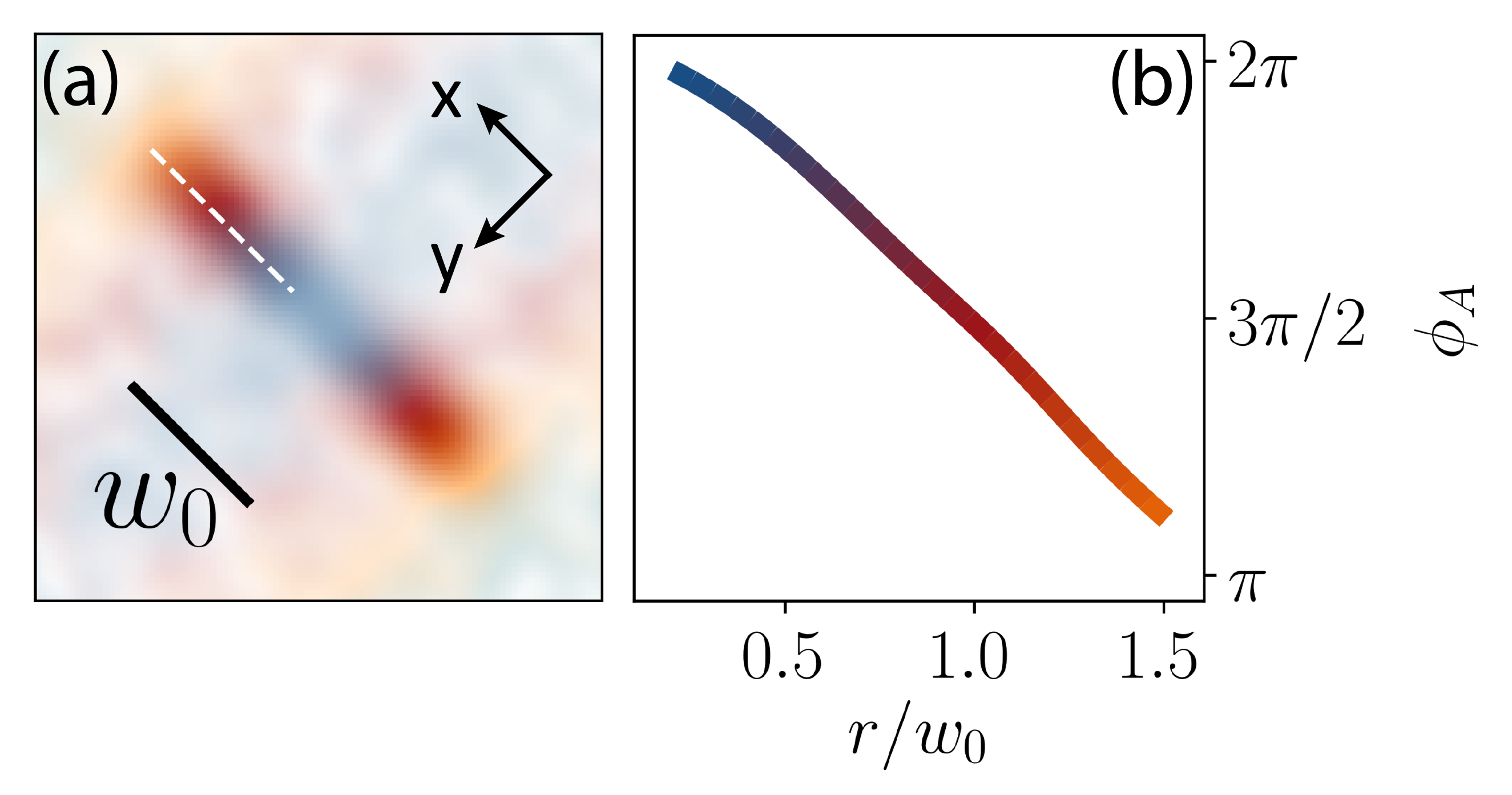}
\caption{(a) Reconstructed cavity superradiance from a BEC with length on the order of the cavity waist $w_0$. The phase of the atomic density wave varies smoothly from one side of a node to the other. Note that the gas is positioned to only one side of the cavity axis. (b) Line-cut of the phase winding as indicated by white dashed line in panel (a). $r$ denotes the distance from the cavity axis. }
\label{largebec}
\end{figure}

We report the direct observation of the  smooth evolution of the density wave phase versus transverse position. We use a BEC with a width sufficiently large  to span two antinodes in the nonlocal interaction and observe a smooth phase evolution of the density waves within the BEC.  That is,  we do not observe special $\mathbb{U}(1)$-symmetry points at which there is a phase freedom because of the expected symmetry breaking.  Rather, the phase smoothly evolves from 0 to $\pi$ as shown in Fig.~\ref{largebec}.  

To observe this, we prepare a BEC that is elongated along one direction using a dithered optical tweezer beam to make an elongated trap. The mean position of the BEC is placed away from the center of the cavity transverse plane. Figure~\ref{largebec} shows the cavity field emission above the self-organization transition threshold. As mentioned above, we observe a smooth phase variation from $0$ to $\pi$ across the BEC straddling the two antinodes of the cavity emission that arises from the nonlocal interaction term. 

While this directly indicates that the phase of the density wave formed in the self-organization transition is no longer restricted to be the same in the entire atomic gas, as was predicted in Ref.~\cite{VaidyaPRX}, we do not observe any shot-to-shot fluctuations in this phase evolution pattern.  The phase evolution is deterministic, by which we mean that the phase winds with the same pattern from experimental shot-to-shot and is predictable by accounting for the effects discussed in the previous section.  That is, there is no fluctuation of the phase or change of the winding orientation indicative of an underlying $\mathbb{U}(1)$-symmetry at these special radii.  By contrast, such fluctuations in the self-organized density wave are expected in a multimode cavity of the ideal concentric configuration~\cite{Gopalakrishnan2009,Gopalakrishnan2010}.  Those works showed how such fluctuations of a  density wave primarily along the cavity axis could lead to a quantum Brazavoskii transition in transversely pumped concentric multimode cavity QED systems.  That is, such fluctuations could drive the system from a Dicke-like superradiant, self-ordering phase transition into a Brazovskii-like superradiant, self-ordering transition, as discussed in Refs.~\cite{Gopalakrishnan2009,Gopalakrishnan2010}. (The Brazovskii transition~\cite{Brazovskii:wr,Hohenberg:1995gm} is a weakly first-order, fluctuation-induced transition that would result in a superfluid smectic---`supersmectic'---state.  The Dicke transition, by contrast, is a second-order mean-field transition to a supersolid state~\cite{Leonard:2017wx}.)

The deterministic phase variation we observe in Fig.~\ref{largebec} illustrates that the preferred density wave phase is different at different locations, but there are no fluctuations in this nonideal confocal cavity.  Moreover, such fluctuations in the \textit{longitudinal} direction are not likely to be possible in even an ideal confocal cavity due to the long Rayleigh length of the modes in such a cavity.  In summary, various corrections to  $\mathcal{D}^{+}_{\mrm{nonlocal}}$ in a nonideal confocal cavity lift the expected $\mathbb{U}(1)$-symmetry at all positions in the multimode cavity other than the ones in the midplane.   This symmetry-breaking will hamper one's ability to observe Brazovskii physics in a nonideal confocal cavity pumped with a single field.   

The next section proposes a new pumping scheme that restores this $\mathbb{U}(1)$-symmetry and thus reintroduces the possibility that fluctuations in the phase of the atomic density waves can drive the system to a supersmectic state via a quantum Brazovskii transition.

\section{Proposal for restoration of $\mathbb{U}(1)$-symmetry}\label{restoration}

The aforementioned rigidity in the allowed phase of the density wave is a direct result of the nonlocal interaction. We now propose a pumping scheme, involving two degenerate resonances in a confocal cavity, for eliminating the $\mathcal{D}^{+}_{\mrm{nonlocal}}$ term from the cavity-mediated interaction.  This would restore the full phase freedom in the atomic density wave, perhaps allowing Brazovskii physics to be explored in a confocal cavity, even a nonideal one.

We suggest adding a second transverse pump at a frequency near a confocal resonance one FSR away from the first pump. As noted in Eq.~\eqref{totalLightField}, the phase variation of different transverse modes depends on the longitudinal index. For a resonance 1 FSR away, the longitudinal index $Q$ changes by 1. Thus, in this newly driven set of degenerate mode family resonances, the $\mu$-independent part of the slowly varying phase changes from $\Theta_{Q_0} \to \Theta_{Q_0} + \pi/2$.  Since the atomic density wave offset $\delta$ can only be adjusted to match one family, we find that for this second mode family, computing the overlap factors as in Eq.~\eqref{overlapfator} gives
\begin{equation}
\mathcal{O}_\mu^{\prime\sigma}= 
\begin{cases} -\sin{[n_\mu \theta_0(z) ]} & \sigma=c \\ \cos{[n_\mu \theta_0(z) ]} & \sigma=s\end{cases}.
\end{equation}
Since these are two nearby resonances within the same cavity, it is reasonable to assume that they have the same coupling $g_0$ and residual mode splittings $\epsilon$. Thus, if we choose the two pumps to have the same effective detuning $\Delta_{Q_0}$ and the same pumping strength $\Omega^2/\Delta^2_a$, the relevant total cavity-mediated interaction $\mathcal{T}$ is simply the sum of the contribution from individual resonances
\begin{align}
\mathcal{T}_{\sigma \tau} &= \Delta_{Q_0}\sum_{\mu}\frac{\Xi_{\mu}(\mbf{r})\Xi_{\mu}(\mbf{r}^\prime)}{\Delta_{\mu}+i\kappa}\left( \mathcal{O}_{\mu}^\sigma \mathcal{O}_{\mu}^{\tau} + \mathcal{O}_{\mu}^{\prime\sigma} \mathcal{O}_{\mu}^{\prime \tau} \right) \mathcal{S}^{+}_{\mu} \nonumber \\
 &\propto \mathbb{1} \mathcal{G}^{+}(\mbf{r},\mbf{r}^\prime,0).
\end{align}
This identity holds because the sign change in $\mathcal{O}_\mu^{\prime c}\mathcal{O}_\mu^{\prime s}$ versus $\mathcal{O}_\mu^{c}\mathcal{O}_\mu^{s}$ cancels the off diagonal components. Therefore, the combined interaction $\mathcal{T}$ is now diagonal in the basis of $\psi_{c,s}$ and the system  exhibits a $\mathbb{U}(1)$-symmetry in the entirety of the cavity's transverse plane. We note that this result is general and does not rely on an ideal cavity or the atoms being at special radii in the midplane of the cavity. Combined with a short-range local interaction $\mathcal{G}^{+}(\mbf{r},\mbf{r}^\prime,0)$, observation of fluctuations in atomic density wave in a large BEC should be possible, which may in turn lead to a Brazovskii transition.  This will be the subject of future work.

\section{Cavity mediated interaction in three dimensions}\label{threeDimensions}

In the discussion so far, we have assumed the cavity couples to atomic density waves.  Under this assumption, combined with the restriction of atoms to a thin cloud with $w_z \ll z_R$, we could ignore the $z$ dependence of the cavity-mediated interaction, and replace it with the matrix structure in terms of sine and cosine quadratures of atomic density waves.  In order to be able to apply the results discussed here to spin models (where atoms could in principle be placed at arbitrary positions $\vec{r}, z$), such as recently realized in a BEC system~\cite{kroeze2018spinor}, it is useful to also present the cavity-mediated interaction with its $z$-dependence included.  That is, we allow for interactions between atoms at distinct longitudinal coordinates $z$ and $z^\prime$. We thus consider:
\begin{equation}
    \mathcal{D}^+_{\text{3D}}(\vec r, \vec r^\prime, z,z^\prime)
    =
    \displaystyle\sum_{\mu}  \frac{\Phi_{\mu}(\mbf{r},z)\Phi_{\mu}(\mbf{r}^\prime,z^\prime) }{1 + \tilde{\epsilon} n_\mu + i \tilde{\kappa}}
    \mathcal{S}_\mu^+,
\end{equation}
where $\Phi_{\mu}(\mbf{r},z)=\Xi_\mu(\mbf{r}) \cos[ k_rz - \Theta_{Q_0}(z) - n_\mu \theta_0(z)]$ is the full mode function.  One may see that this takes the form:
\begin{multline}
    \mathcal{D}^+_{\text{3D}}(\vec r, \vec r^\prime, z,z^\prime)
    =
    \frac{1}{4} 
    \sum_{\tau,\tau^\prime = \pm}
    \mathcal{G}\left(\vec r, \vec r^\prime, - i [\tau \theta_0(z) + \tau^\prime \theta_0(z^\prime)]\right) 
    \\\times 
    e^{i k_r (\tau z+ \tau^\prime z^\prime) 
    - i [\tau \Theta_{Q_0}(z) + \tau^\prime \Theta_{Q_0}(z^\prime)
    ]}.
\end{multline}
This expression represents the most general cavity-mediated interaction in a near-confocal cavity.

Moreover, we  can now transform this expression into the form used in the companion paper~\cite{GouyPRL2018}. To do so, we should consider this expression under the special condition that we work near the midpoint of the cavity, so that the $z$-separation between atoms is small compared to the Rayleigh range. We may then neglect the $z$-dependence of $\theta_0$ and $\Theta_{Q_0}$, and use $\mathcal{G}(\vec{r},\vec{r}^\prime, -i \pi/2)=\mathcal{G}(\vec{r},\vec{r}^\prime, i \pi/2)$ to write:
\begin{multline}
    \mathcal{D}^+_{\text{3D}}(\vec r, \vec r^\prime, z,z^\prime)
    =
    \frac{1}{2} 
    \Big[
    \mathcal{G}\left(\vec r, \vec r^\prime, 0\right) 
     \cos(k_r(z-z^\prime))
     \\+
    \mathcal{G}\left(\vec r, \vec r^\prime, - 2 i \theta_0 \right) 
    \cos( k_r (z+ z^\prime) - 2 \Theta_{Q_0})
    \Big].
\end{multline}
For even cavities, $\Theta_{Q_0}$ is an integer multiple of $\pi$, and thus we can write:
\begin{multline}
    \mathcal{D}^+_{\text{3D}}(\vec r, \vec r^\prime, z,z^\prime)
    =
    \frac{1}{2} 
    \Big[
    \mathcal{G}_{\text{local}}\left(\vec r, \vec r^\prime\right) 
     \cos[k_r(z-z^\prime)]
     \\ \pm
    \mathcal{G}_{\text{nonlocal}}\left(\vec r, \vec r^\prime\right) 
    \cos[ k_r (z+ z^\prime) ]
    \Big].
\end{multline}
This expression matches the form of the interaction introduced in the companion paper~\cite{GouyPRL2018}.  Integrating this over atomic density waves naturally recovers the interaction matrix discussed in Sec.~\ref{selforg}, while for spin degrees of freedom, this expression can be directly applied, depending on the locations of trapped spinful atoms.

\section{Beyond confocal cavities}
\label{beyond-confocal}

\begin{figure}
    \centering
    \includegraphics[width = 0.49\textwidth]{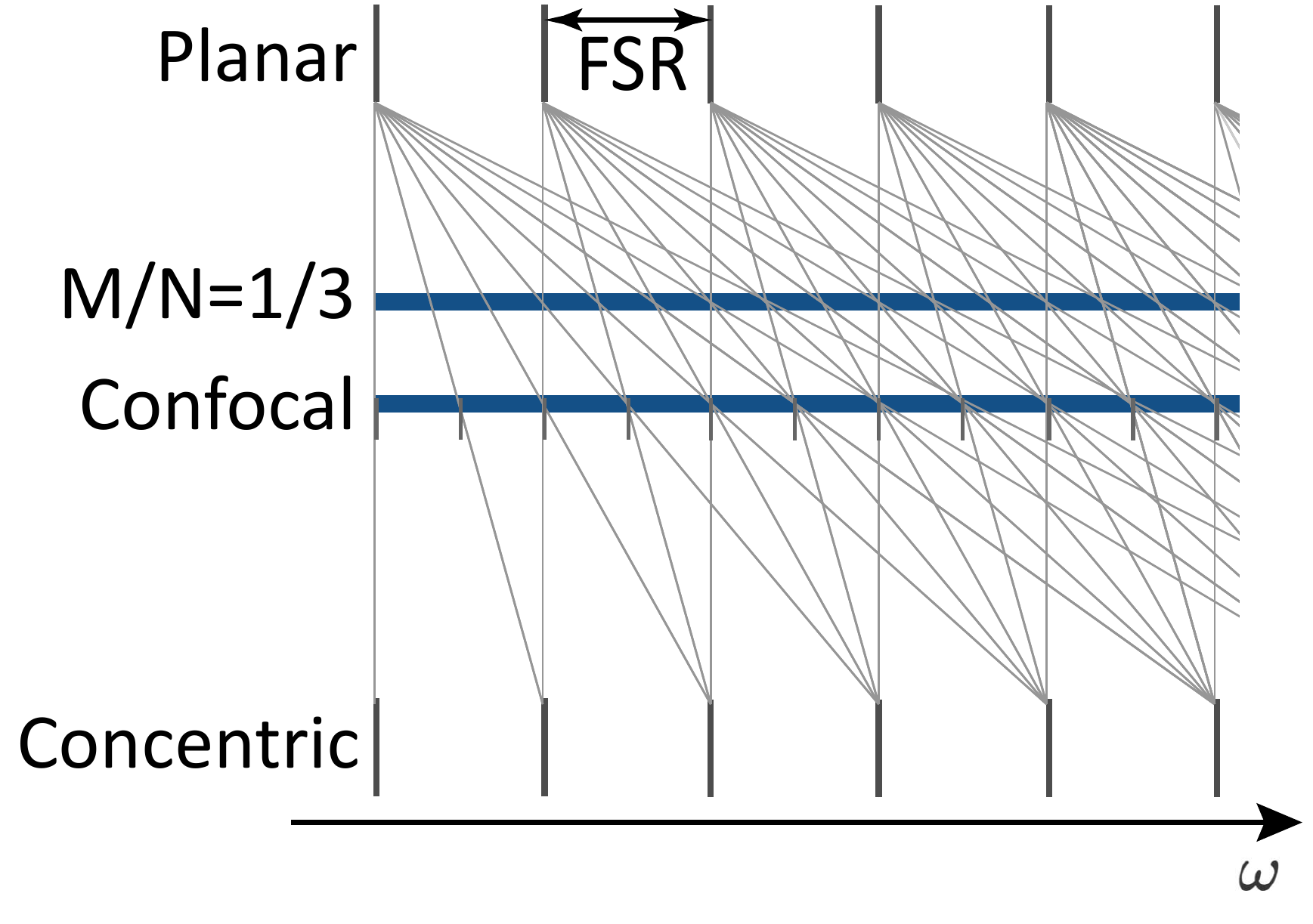}
    \caption{Diagram illustrating evolution of cavity frequencies versus changing cavity length. Black  vertical lines at the top of the panel indicate different longitudinal resonances, separated by one free spectral range (FSR).  As one moves from the planar limit (top, $L \ll R$) to the concentric limit (bottom, $L=2R$), the different transverse modes within a given mode family split and cross with those from higher longitudinal families.  Some degenerate configurations are indicated by horizontal blue lines.}
    \label{fig:higher_resonances}
\end{figure}

All the results so far are based on cavity configurations near the confocal geometrical configuration.  However, there exist a much wider set of Fabry-P\'{e}rot cavity configurations with resonances at which multiple modes become degenerate.  To describe these cases, let us note that we can consider all degeneracies as arising from the way the frequencies of different transverse modes shift and cross as one varies the length of the cavity from a short (nearly planar) configuration of $L\ll R$, past the confocal point of $L=R$ and toward the concentric limit of $L=2R$.  This is illustrated in Fig.~\ref{fig:higher_resonances}.
In the following, we first review the basic physics of higher-order degeneracies, generalizing the  discussion of Ref.~\cite{siegman1986lasers}.  We provide the following summary here for completeness.

\subsection{Generalized degeneracy conditions}

The cavity lengths at which degeneracies occur can be understood by considering the frequency at which the fundamental (i.e., TEM$_{00}$ mode) of one family is crossed by the $N$th transverse modes of a family that is $M$-FSRs away.  That is to say, the $n_\mu=0$ mode of one family (with longitudinal mode number $Q_0$) is degenerate with the modes $n_\mu = N$ from a different family with longitudinal mode number $Q_0-M$.  We refer to these as $M/N$ resonances.  For example, the confocal situation corresponds to $M/N=1/2$, and concentric to $M/N=1$. Due to the linearity of the frequency shifts with mode index, we may note that at the $M/N$ degeneracy, as well as the $n_\mu=0$ mode of one family being degenerate with the $n_\mu=N$ mode of the family $M$-FSR away, it is also degenerate with the $n_\mu=2N$th mode of the family $2M$ away, and so on. Therefore, the cavity becomes highly degenerate.  Moreover, the 1st-order transverse modes of a given family will be degenerate with the $N+1$th modes of the family $M$ away, etc.  Thus, there are $N$ distinct types of degenerate points, generalizing the two odd/even sets of degeneracy families of the confocal case.

The condition on cavity length and mirror curvature required for such a degeneracy can be found by requiring equal wavevectors between pairs of modes indexed as described above.  Using the mode functions as written in Eq.~\eqref{totalLightField}, and in particular, the phase $\theta_{\mu,Q}(z)$ written in Eq.~\eqref{eq:lightfieldphase}, we find that to match the boundary conditions that cavity light vanishes on-axis at the mirrors at $z=\pm L/2$, we require that
\begin{align}
    -k_{\mu,Q} \frac{L}{2} - \theta_{\mu,Q}\left(-L/2\right)
    &= \frac{\pi}{2},
    \label{bc-theta-a}
    \\
    k_{\mu,Q} \frac{L}{2} - \theta_{\mu,Q}\left(\phantom{-}L/2\right)
    &= \frac{\pi}{2} + Q \pi.
    \label{bc-theta-b}
\end{align}
where $k_{\mu,Q}$ is the wavevector associated with the mode with transverse index $\mu$ and longitudinal modenumber $Q$.  Note that as per its definition, $Q$ counts the phase difference between the mirrors, thus labelling the longitudinal mode number.
In addition, to match the boundary condition across the transverse plane of the mirrors, we require that the radius of curvature $R(z)$ at the mirror locations $z=\pm L/2$ should match the mirror curvature $R$.  

To solve Eqs.~\eqref{bc-theta-a} and~\eqref{bc-theta-b}, we take sums and differences.  Using the definition of $\theta_{\mu,Q}(z)$ in Eq.~\eqref{eq:lightfieldphase}, these yield the following equations:
\begin{align}
    \label{eq:define-k}
    k_{\mu,Q} L &= Q \pi + 2 \psi(L/2) (1+ n_\mu), 
    \\
    \label{eq:define-xi}
    2 \xi_{\mu,Q} &=(Q+1) \pi + 2 n_\mu \psi(L/2).
\end{align}
The first of these equations is the condition to solve for degenerate points.  Specifically, our $M/N$ resonance means that the values of $k_{\mu,Q}$ should be equal for $(n_\mu,Q)$ and $(n_\mu+N, Q-M)$. The condition for this to occur is thus $M \pi = 2 N \psi(L/2)$.  Using the definition of $\psi(z)$, we may regard this condition as defining $z_R$ in terms of $L/2$ to ensure resonance.  We can then use this, combined with the definition of $R(z)$, to find the radius of curvature at the mirrors.  We thus find the $M/N$ resonance corresponds to:
\begin{displaymath}
  z_R = \frac{L}{2} \cot\left( \frac{M \pi}{2 N} \right),
  \qquad
  R = \frac{L}{2} \text{cosec}^2\left(\frac{M \pi}{2 N} \right).
\end{displaymath}
From this condition we may note that even with a small range of tunability of cavity length $L$, one can nonetheless realize high-order resonances by finding irreducible fractions $M/N$ near the confocal point of $M/N=1/2$.

We may note that since the degeneracy is between transverse modes with total index $n_\mu$ separated by $N$, there will be $N$ separate degeneracies, corresponding to $\eta = n_\mu \mod N$, with $\eta \in \{0 \ldots N-1\}$.  Each of these families has an offset cosine along the longitudinal direction. To see this, one may note that from the degeneracy condition between modes $(n_\mu,Q)$ and $(n_\mu+N, Q-M)$, we may thus label $n_\mu = \eta + P N, Q=Q_0-P M$ for integer $P$, where $Q_0$ is the longitudinal mode number of the lowest transverse mode in the given family.  As a result $\xi_{\mu,Q} = \xi_{Q_0} = (\pi/2)(Q_0 +1+ \eta M/N) $ is constant within a given degenerate family.  We also see that as with the confocal case, families separated by one FSR differ in phase by $\pi/2$, giving an associated offset of the atomic density waves.   We may note that while for the confocal resonance, different mode families had orthogonal longitudinal dependence, this is not true for general $M/N$.

\subsection{Generalized interaction matrix}

Having found the resonances, we may now consider one near-degenerate family with $k_{\mu,Q}=k_r$ and find the effective interaction matrix.
Focusing on a single degenerate family, we consider atomic density waves of the form $\cos(k_r z +\delta), \sin(k_r z +\delta)$
with a phase offset $\delta=\Theta_{Q_0}(z_0)$ chosen to match the given family.  One may show that the matrix describing the cavity-mediated matrix of Green's functions takes exactly the same form as in Eq.~\eqref{eq:kernel}, but with modified overlap functions and a modified factor $\mathcal{S}_\mu$.  Specifically, the angle $\theta_0(z_0)$ appearing in $\mathcal{O}^\sigma_\mu$ is now given by $\theta_0(z_0) = \psi(z_0) + \psi(L/2) = \arctan(z_0/z_R) + M \pi/2 N$, and the factor restricting the mode summation should now be:
\begin{displaymath}
  \mathcal{S}^\eta_\mu = \frac{1}{N}\sum_{s=0}^{N-1}
  \exp(i 2 \pi s(n_\mu - \eta)/N).
\end{displaymath}
This gives $1$ for $n_\mu = \eta \mod N$, and zero otherwise.  The label $\eta$ thus replaces the $\pm$ label for even/odd modes. We can then similarly define  
\begin{equation}
    \mathcal{G}^{\eta}(\vec r, \vec r^\prime , \varphi)=
    \frac{1}{N}
    \sum_{s=0}^{N-1} 
    e^{-i 2 \pi {s \eta}/{N} }
    \mathcal{G}(\mbf{r},\mbf{r}^\prime,\varphi-i 2s \frac{\pi}{N}).
\end{equation}
Using the above, we can then calculate the interaction matrix, the equivalent of Eq.~\eqref{confocalkernel}; we find:
\begin{align}
2\mathcal{D}^{\eta} (\mbf{r},\mbf{r}^\prime) &=  
\mathbb{1} \mathcal{G}^\eta(\mbf{r},\mbf{r}^\prime,0) 
\nonumber\\ &+ 
\frac{\sigma^{z}}{2} \left[ 
\mathcal{G}^\eta (\mbf{r},\mbf{r}^\prime,-2 i \theta_0) + \mathcal{G}^\eta (\mbf{r},\mbf{r}^\prime,2 i \theta_0) \right]
\nonumber\\&+
\frac{\sigma^{x}}{2i} \left[ 
\mathcal{G}^\eta (\mbf{r},\mbf{r}^\prime,-2 i \theta_0) - \mathcal{G}^\eta (\mbf{r},\mbf{r}^\prime,2 i \theta_0) \right],
\label{generalconfocalkernel}
\end{align}
which is identical in structure to the confocal case, up to the replacement of $\mathcal{G}^+ \to \mathcal{G}^\eta$ and the modified value of $\theta_0$.  One may also verify that by taking $M/N\to 1/2$ and using $\eta=0,1$ for the $\pm$ families, we recover the confocal results.

In the confocal case, neglecting effects of finite cloud size, we noted that at the cavity midplane, $z_0=0$, and the coupling between sine and cosine density waves vanished at all radii.  The same applies here for $\eta=0$ (analogous to the even modes considered before).  In fact, we can extend this to all $\eta$ if we consider an $\eta$-dependent shift to the phase offset $\delta$.  Specifically, we choose the offset such that the overlap factors become
\begin{equation}
    \mathcal{O}_{\mu}^\sigma=
    \begin{cases} 
    \cos(n_\mu \theta_0 - \pi M \eta/2N) & \sigma=c 
    \\ 
    \sin(n_\mu \theta_0 - \pi M \eta/2N) & \sigma=s
    \end{cases}.
\end{equation}
With this, we find that at the cavity midplane (where $\theta_0 = M \pi/2 N$), the remaining kernel takes the form:
\begin{multline}
2\mathcal{D}^{\eta} (\mbf{r},\mbf{r}^\prime) =  
\mathbb{1} \mathcal{G}^\eta(\mbf{r},\mbf{r}^\prime,0) 
\\ + 
\sigma^{z} e^{-i \pi M \eta/N}
\mathcal{G}^\eta (\mbf{r},\mbf{r}^\prime,- i M \pi /N)
\label{centralgeneralconfocalkernel}.
\end{multline}

In terms of the spatial structure of these functions, there is one notable differences between the confocal case and the higher-order resonances.  This is that in the confocal case, $N=2$, and at the cavity midplane, the angle $\theta_0=\pi/4$ led to an especially simple form of the nonlocal interaction, $\cos(2 \vec{r} \cdot \vec{r}^\prime/w_0^2)$. For the higher-order resonances, no such simplification occurs, and thus the nonlocal interaction takes a form involving a sum of terms of the more general form $\cos(A + B \vec{r}\cdot\vec{r}^\prime / w_0^2 + C(\vec{r}^2+\vec{r}^{\prime 2})/w_0^2)$.   Thus, for such a higher resonance, the output light coming from a single spot involves not only a set of parallel fringes (contours of constant $\vec{r}\cdot\vec{r}^\prime$, but also a set of circular fringes around the location of the atom cloud.  Whether such interactions could be used for engineering interesting cavity-mediated spin-spin interactions is the subject of future work. 

\section{Conclusions and Outlook}

We have shown that the structure of Gaussian modes in a near-confocal multimode cavity leads to cavity-mediated atom-atom interactions that can favor distinct patterns of atomic density waves as one moves transversely across the cavity.  We find, moreover, that there can exist specific radii where there is phase freedom for the atomic density wave.  For a perfectly confocal cavity and small-sized BECs, such radii exist at all longitudinal distances along the cavity.  Including effects of finite nonconfocality, the emergent symmetry exists on the midplane of the cavity, while elsewhere we have symmetry breaking.  Nonetheless, the symmetry-broken state still shows interesting phase evolution of the atomic density wave versus radius, smoothly evolving between sine and cosine density waves as the atomic gas is moved across the transverse plane.   Such behavior matches that which we report experimentally.

The results presented here provide the foundation to further control  the cavity-mediated interaction among intracavity atoms and atomic spins.  In particular, we show that a configuration involving two pump laser frequencies can induce a full $\mathbb{U}(1)$-symmetry, even accounting for imperfections of the real cavity.  This provides a route to exploring liquid crystalline order~\cite{Gopalakrishnan2009,Gopalakrishnan2010} in experimentally accessible conditions with a confocal cavity. At the same time, we showed how other, more complex, interactions can be engineered, by using higher-order cavity resonances.  In particular, since resonances are labelled by irreducible fractions
$M/N$,  high-order resonances can be realized without requiring a significant change of cavity length.  These higher-order resonances provide a further tool in realizing tunable cavity-mediated interactions.

\begin{acknowledgements}
We acknowledge funding support from the Army Research Office, the National Science Foundation under Grant No.~CCF-1640075, and the Semiconductor Research Corporation under Grant No.~2016-EP-2693-C.  J.~K. acknowledges support from SU2P.
\end{acknowledgements}

\appendix

\section{Finite BEC extent along cavity axis}

One additional effect that may be considered as a source of breaking of the $\mathbb{U}(1)$-symmetry is the finite extent of the atom gas along the cavity axis.  This corresponds to relaxing our simplifying assumption that the phase of cavity modes is constant over the extent of the atomic gas, and can result in a more complicated overlap factor in Eq.~\eqref{overlapfator}.  However, as seen below, this effect on its own does not to lead to a smooth phase evolution.

To account for this effect, we must return to the definition of the overlap factor to include variation of the Gouy phase in the integral over $z$ that appears in this factor. That is,  we cannot write this effect as a single integral over $z$ at the end of the calculation.
From Eq.~\eqref{zphase}, the phase variation along the cavity axis direction is $\theta_0 (z) = \pi/4 + \mrm{arctan}\left(z/z_R\right) 
$. We may expand around the midpoint of the atom cloud, $z_0$, and linearize to write
$\theta_0 (z) \approx \theta_0(z_0) + \theta^\prime_{0}(z_0) (z-z_0)$. The modified overlap factor should then include an integral over the density profile $|Z(z-z_0)|^2$ of the atoms along the $z$ direction
\begin{equation}
\label{gradientoverlap}
\mathcal{O}_\mu^\sigma= \int dz |Z(z-z_0)|^2 \begin{cases} \cos{[n_\mu \theta_0(z) ]} & \sigma=c \\ \sin{[n_\mu \theta_0(z) ]} & \sigma=s\end{cases}.
\end{equation}
Assuming a Gaussian density profile 
\be
|Z(z-z_0)|^2 = \frac{1}{\sqrt{2 \pi w_z^2}} \exp\left(-\frac{(z-z_0)^2}{2 w_z^2}\right),
\ee
the result of the integral is given by
\be
\mathcal{O}_\mu^\sigma= \begin{cases} \cos{[n_\mu \theta_0(z_0) ]} \mrm{exp}\left[ -\frac{n^2_{\mu} (w_z \theta_0^\prime(z_0))^2}{2}\right] & \sigma=c \\ \sin{[n_\mu \theta_0(z_0) ]} \mrm{exp}\left[ -\frac{n^2_{\mu} (w_z \theta_0^\prime(z_0))^2}{2}\right] & \sigma=s \end{cases}.
\ee
Because of the appearance of the $n^2_{\mu}$ factor in the exponent, we can no longer apply Eq.~\eqref{harmonicgreen} to evaluate the sum in Eq.~\eqref{kernelS}. We can nonetheless see that the off-diagonal term still vanishes for $\theta(z_0) = \pi/4$. Thus, once again, we see the variation of the cavity mode phase has no effect on the symmetry breaking for atoms located at the midplane of the cavity.  The presence of the $n^2_{\mu}$ factor in the exponent also means we can no longer find a simple expression for the effect of this term away from the cavity midplane.  We leave additional  explorations of this effect to future work. 

\pagebreak


%

\end{document}